%% file: main.tex
\documentclass[sigconf]{acmart}
\setcopyright{acmcopyright} 
\acmConference[Anonymous Submission to ACM CCS]{Conference on Computer and Communications Security}
\acmYear{2026}

\usepackage{booktabs} 
\usepackage{graphicx}
\usepackage{epstopdf}
\usepackage{makecell}
\usepackage[flushleft]{threeparttable}
\usepackage{listings}
\usepackage{xcolor}
\usepackage{balance}
\usepackage{amsmath}

\usepackage{tabu}
\usepackage{xspace}
\usepackage{multirow}

\usepackage{nimbusmononarrow}
\usepackage[T1]{fontenc}

\usepackage{textcomp,booktabs}
\usepackage{colortbl}
\usepackage{marvosym}
\usepackage{algorithm}
\usepackage{algorithmicx}
\usepackage{algpseudocode}

\usepackage{pifont}
\usepackage{listings}
\usepackage{cleveref}
\usepackage{enumitem}
\usepackage{microtype}
\usepackage{ulem}
\usepackage{stfloats}
\usepackage{fancyhdr}
\usepackage{balance}
\usepackage{tabularray} 
\usepackage{seqsplit}
\usepackage{xspace}
\usepackage{fontawesome5}
\usepackage{marvosym}
\usepackage{multirow}
\usepackage{color,xcolor,colortbl}
\usepackage{soul}
\usepackage{framed}

\usepackage{hyperref}
\hypersetup{
    colorlinks = true,
    urlcolor = black,
    linkcolor = black,
    citecolor = black
}

\usepackage[all]{nowidow}

\usepackage{xurl} 
\usepackage{hyperref} 

\makeatletter
\newcommand{\thickhline}{%
	\noalign {\ifnum 0=`}\fi \hrule height 0.8pt
	\futurelet \reserved@a \@xhline
}
\newcolumntype{"}{@{\hskip\tabcolsep\vrule width 0.8pt\hskip\tabcolsep}}
\newcolumntype{*}{!{\vrule width 0.8pt}}
\makeatother

\newcommand{\squishlist}{
\begin{itemize}[noitemsep,nolistsep,leftmargin=\parindent]
  \setlength{\itemsep}{-0pt}
  \setlength{\parskip}{0pt}
}
\newcommand{\squishend}{
  \end{itemize}
}

\newcommand{\ignore}[1]{}

\newcommand{\sysname}{{\sc MiniAuth}\xspace}

\definecolor{mygreen}{rgb}{0,0.6,0}
\definecolor{mymauve}{rgb}{0.58,0,0.82}
\definecolor{ashgrey}{rgb}{0.7, 0.75, 0.71}
\definecolor{mygrey}{rgb}{0.85, 0.85, 0.85}

\definecolor{codegreen}{rgb}{0,0.6,0}
\definecolor{codegray}{rgb}{0.5,0.5,0.5}
\definecolor{codepurple}{rgb}{0.58,0,0.82}
\definecolor{backcolour}{rgb}{0.95,0.95,0.95}

\definecolor{mygray}{gray}{.9}

\lstdefinestyle{mystyle}{
	backgroundcolor=\color{backcolour},
	commentstyle=\color{codegreen},
	keywordstyle=\color{magenta},
	numberstyle=\ttfamily\small\color{codegray},
	stringstyle=\color{codepurple},
	mathescape=true,
	xleftmargin=8pt,
	xrightmargin=5pt,
	basicstyle=\ttfamily\small,
	breakatwhitespace=false,
	breaklines=true,
	captionpos=b,
	numbers=left,
	numbersep=5pt,
	showspaces=false,
	showstringspaces=false,
	showtabs=false,
	tabsize=2,
	frame=single,
	moredelim=**[is][\color{red}]{@}{@},
}

\definecolor{mygreen}{rgb}{0,0.6,0}
\definecolor{mymauve}{rgb}{0.58,0,0.82}
\definecolor{ashgrey}{rgb}{0.7, 0.75, 0.71}
\definecolor{mygrey}{rgb}{0.85, 0.85, 0.85}
\definecolor{lightSkyBlue}{RGB}{135,206,250}
\definecolor{codegreen}{rgb}{0,0.6,0}
\definecolor{codegray}{rgb}{0.5,0.5,0.5}
\definecolor{codepurple}{rgb}{0.58,0,0.82}
\definecolor{backcolour}{rgb}{0.95,0.95,0.95}
\definecolor{mygray}{gray}{.9}
\sethlcolor{lightSkyBlue}
\lstdefinestyle{mystyle}{
	backgroundcolor=\color{backcolour},
	commentstyle=\color{codegreen},
	keywordstyle=\color{magenta},
	numberstyle=\ttfamily\footnotesize\color{codegray},
	stringstyle=\color{codepurple},
	mathescape=true,
	xleftmargin=10pt,
	xrightmargin=5pt,
	basicstyle=\ttfamily\small,
	breakatwhitespace=false,
	breaklines=true,
	captionpos=b,
	numbers=left,
	numbersep=5pt,
	showspaces=false,
	showstringspaces=false,
	showtabs=false,
	tabsize=2,
	frame=single,
	moredelim=**[is][\color{red}]{@}{@},
}

\lstdefinelanguage{JavaScript}{
    keywords={typeof, new, true, false, catch, function, return, null, catch, switch, var, if, in, while, do, else, case, break, this,wx,navigateTo,request,login,reLaunch,setData,showToast,url,my,swan,JSON,stringify,setNavigationBarTitle,parse,getStorageSync,swan},
    keywordstyle=\color{blue}\bfseries,
    ndkeywords={},
    ndkeywordstyle=\color{darkgray}\ttfamily,
    identifierstyle=\color{black},
    sensitive=false,
    comment=[l]{\#},
    morecomment=[s]{/*}{*/},
    commentstyle=\color{magenta}\ttfamily,
    moredelim=[is][\textcolor{qinglv}]{\%\%}{\%\%},
    moredelim=**[is][\color{blue}]{\$\$}{\$\$},
    escapeinside={(*@}{@*)}, 
}

\usepackage{tcolorbox}
\tcbuselibrary{breakable, skins}
\usepackage{caption}

\tcbset{%
	label begin/.style={label={#1}},
	label end/.style={after upper=\label{#1}}
}

\newtcolorbox[%
auto counter]{mybox}[2][]{%
	enhanced jigsaw,
	breakable,
	#1}

\usepackage{tikz}
\newcommand*{\circled}[1]{\lower.7ex\hbox{\tikz\draw (0pt, 0pt)%
		circle (.4em) node {\makebox[0.25 em][c]{\small #1}};}}

\newcommand{\paragraphNew}[1]{\vspace{3pt}\noindent{\bf{#1}.}}

\begin{document}

\title[Mini-Programs, Mega-Problems: Unveiling OAuth-based Authentication Misuses in Mini-Programs via Dynamic Analysis]{Mini-Programs, Mega-Problems: Unveiling OAuth-based Authentication Misuses in Mini-Programs via Dynamic Analysis}






\author{Zidong Zhang}
\authornote{Both authors contributed equally to this research.}
\affiliation{%
  \institution{Simon Fraser University; QI-ANXIN Research Institute}
  \city{Burnaby}
  \country{Canada}
}
\email{zza323@sfu.ca}

\author{Zhentao Xie}
\authornotemark[1]
\affiliation{%
  \institution{Shandong University; The Chinese University of Hong Kong}
  \city{Qingdao}
  \country{China}
}
\email{fxizenta@mail.sdu.edu.cn}

\author{Lingyun Ying}
\authornote{Corresponding authors.}
\affiliation{%
  \institution{QI-ANXIN Technology Research Institute}
  \city{Beijing}
  \country{China}
}
\email{yinglingyun@qianxin.com}

\author{Qinsheng Hou}
\affiliation{%
  \institution{School of Computer Science, Shanghai Jiao Tong University}
  \city{Shanghai}
  \country{China}
}
\email{houqinsheng@sjtu.edu.cn}

\author{Yacong Gu}
\affiliation{%
  \institution{Tsinghua University; Tsinghua University-QI-ANXIN Group JCNS}
   \city{Beijing}
  \country{China}
}
\email{guyacong@tsinghua.edu.cn}

\author{Wenrui Diao}
\authornotemark[2]
\affiliation{%
  \institution{School of Cyber Science and Technology, Shandong University}
  \city{Qingdao}
  \country{China}
}
\email{diaowenrui@link.cuhk.edu.hk}

\author{Jianliang Wu}
\authornotemark[2]
\affiliation{%
  \institution{Simon Fraser University}
  \city{Burnaby}
  \country{Canada}
}
\email{wujl@sfu.ca}

\renewcommand{\shortauthors}{Zidong Zhang, et al.}



\input{0.abstract}

\maketitle

\input{1.Introduction}
\input{2.Background}

\input{3.Motivation}
\input{4.Challenges}

\input{5.Design}

\input{7.Implementation}

\input{6.Evaluation}

\input{8.Discussion}

\input{9.RelatedWork}
\input{10.Conclusion}
\input{Ethical_OpenScience}

\bibliographystyle{ACM-Reference-Format}
\balance
{\normalem
\bibliography{refs}
}

\input{Appendix}

\end{document}

%% file: 0.abstract.tex
\begin{abstract}


Mini-programs have become a dominant paradigm for lightweight application deployment within super apps such as WeChat. To support seamless integration, super apps provide platform-specific OAuth-based Authentication (OBA) mechanisms for user login. However, improper integration of OBA flows by third-party developers can lead to critical security flaws.

In this paper, we systematize three runtime OBA misuse patterns that arise in mini-program implementations and enable attackers to impersonate victims. To assess their real-world impact, we design and implement \sysname, the first analysis framework to systematically analyze mini-program OBA misuse at scale. \sysname automatically pinpoints the OBA login page of a mini-program, executes the workflow dynamically, and analyzes its runtime behaviors. This enables it to handle obfuscated mini-programs and uncover vulnerabilities within dynamic behaviors that existing approaches cannot detect.

Applying \sysname to 44,273 WeChat and 2,721 Baidu mini-programs, we uncover 1,834 misuse cases, including critical logic flaws that enable client-side identity forgery via exposed credentials and authentication bypass through static or plaintext identifiers. Our cross-platform evaluation further shows that such misuses are not confined to a single ecosystem but consistently appear across different mini-program platforms. We also identify a cryptographic design flaw in Baidu’s OBA APIs that allows brute-forcing of session keys. We responsibly disclosed our findings to the developers and platforms, receiving acknowledgments and assigned CNVD/CNNVD IDs. These results underscore the need for more robust developer guidance and enhanced platform-level safeguards.

\end{abstract}

%% file: 1.introduction.tex
\section{Introduction}
\label{sec:introduction}

Mini-programs, lightweight applications embedded within super apps such as WeChat and Baidu, have emerged as a dominant platform for mobile services, including finance, healthcare, and government portals. To facilitate identity management, these platforms adopt platform-specific OAuth-based Authentication (OBA) mechanisms. These protocols allow mini-programs to authenticate users and retrieve sensitive data, such as phone numbers or user profiles, from the host super app.

Unlike standardized web or native mobile OAuth flows, OBA in mini-programs is tightly coupled with the closed ecosystem of the super app. Developers must rely on platform-specific APIs, such as \texttt{wx.login} and \texttt{code2Session}, to complete authentication. This vendor-locked design increases implementation complexity. Consequently, developers often misimplement these flows, leading to severe security risks like session hijacking, identity forgery, and data leakage.

Existing research on mini-program security has primarily focused on static issues such as hard-coded credentials~\cite{zhang2023don} and permission misuse~\cite{wang2024minichecker,DBLP:journals/tosem/WangLWL0LW24}. However, these approaches are fundamentally limited in detecting OBA misuse. First, static analysis cannot effectively handle the obfuscated mini-programs common in production. Second, static methods fail to capture dynamic OBA vulnerabilities that emerge during \textit{runtime}, such as flawed client-server interactions or insecure handling of sensitive data during network communication.


\begin{figure}[t]
	\centering
	\includegraphics[width=0.8\linewidth]{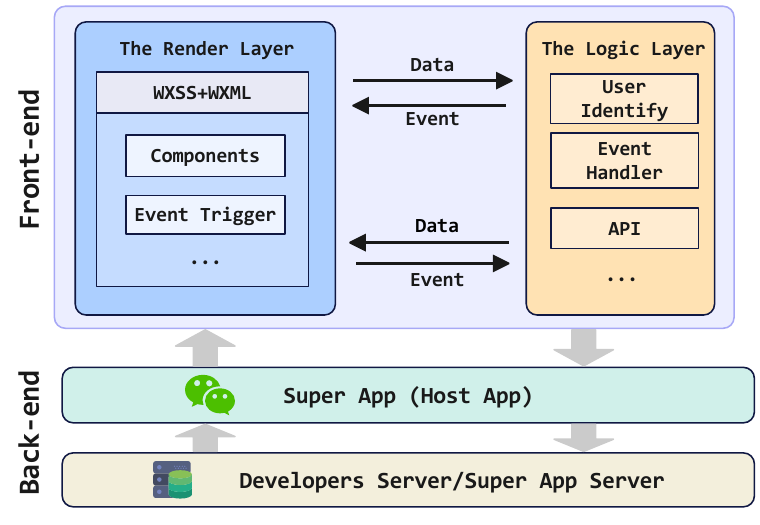}
	\caption{WeChat mini-program architecture.}
 \label{fig:wechat_mini_architecture}
\end{figure}

\paragraphNew{Our Work} In this paper, we identify and summarize three runtime OBA misuse patterns in mini-programs, ranging from client-side identity forgery to the use of static or plaintext identifiers for authentication. These vulnerabilities arise from incorrect implementation or flawed handling of the authentication workflow. Additionally, we reveal a design flaw in a specific platform where the session key is partially leaked through the Initialization Vector (IV) because the IV reuses part of the session key instead of being randomly generated. These issues enable attackers to impersonate victims, log into their accounts, or steal private information.

To systematically uncover these flaws and evaluate their impact, we design and implement \sysname, the first large-scale dynamic analysis framework tailored for detecting mini-program OBA misuses at runtime. \sysname automatically identifies the OBA login page and triggers the workflow to capture dynamic behaviors. Consequently, \sysname can analyze obfuscated mini-programs and uncover runtime-only security issues that are beyond the reach of existing approaches.

We apply \sysname to WeChat and Baidu, the two largest mini-program platforms with 1.385B and 0.704B monthly active users, respectively. We analyze a large dataset of 44,273 and 2,721 mini-programs across these ecosystems. Our analysis reveals widespread OBA misuse issues in the real world. In total, we identify 1,834 misuse cases from 1,688 mini-programs, including 619 programs with multiple vulnerabilities. Alarmingly, several government and healthcare mini-programs are affected, potentially exposing national IDs and personal health records. Our findings indicate that other popular platforms like Alipay and TikTok exhibit consistent insecure practices, suggesting that flawed business logic is common across the industry. Finally, we summarize the root causes and provide suggestions for both developers and platforms to improve OBA security.

\paragraphNew{Contributions} Our contributions are summarized as follows: 
\begin{itemize} 
\item We summarize three OBA runtime misuse patterns and uncover one platform-level design flaw that enables attackers to impersonate users or steal private information.
\item We design and build \sysname, the first dynamic analysis framework that identifies these issues across both obfuscated and non-obfuscated mini-programs using hybrid automation techniques. 
\item We evaluate \sysname on two major platforms with over 46,000 mini-programs and uncover widespread misuses. Our findings have been acknowledged with 11 CNVD/CNNVD~\cite{CNVD,CNNVD} IDs.
\end{itemize}


\paragraphNew{Roadmap} The rest of this paper is organized as follows. Section~\ref{sec:background} provides OBA background. Section~\ref{sec:motivation} details our threat model and misuse taxonomy. Section~\ref{sec:challenges} addresses technical challenges. Sections~\ref{sec:design} and \ref{sec:implementation} describe MINIAUTH , followed by evaluation in Section~\ref{sec:evaluation}. Finally, we discuss implications and limitations in Section~\ref{sec:discussion} , review related work in Section~\ref{sec:related_work} , and conclude in Section~\ref{sec:conclusion}.

%% file: 2.background.tex
\section{Background }
\label{sec:background}

In this section, we introduce the necessary background of the mini-program ecosystem and its OBA mechanism.


\paragraphNew{Mini-program Framework} Mini-programs run within their host super apps, such as WeChat and Baidu. Essentially, these mini-programs are WebView-based applications powered by a runtime engine embedded within the super app.
As shown in \Cref{fig:wechat_mini_architecture}, the mini-program architecture typically consists of two primary components: the \textit{front-end} and the \textit{back-end}. 

The front-end operates within the super app, providing the user interface and handling system-level interactions. It includes a rendering layer, handling the UI and static resources, and a logic layer that processes user inputs and communicates with the back-end.
These layers are managed through a WebView thread and a JavaScript thread, respectively.
Different super apps may implement unique front-end frameworks.
For instance, WeChat designed the HTML-like language WXML~\cite{wxml} for rendering, whereas Baidu uses a customized SWAN~\cite{baiduswan} framework.

The back-end of a mini-program typically includes two components: the super app servers and the mini-program developer’s own server.
The platform provides runtime support and integration with system features, such as payments and messaging.
Meanwhile, most mini-programs with dynamic content or user-specific services rely on dedicated back-end servers for data storage, user management, and other logic, though not all mini-programs require one.


\begin{figure*}[h]
	\centering
	\includegraphics[width=1\linewidth]{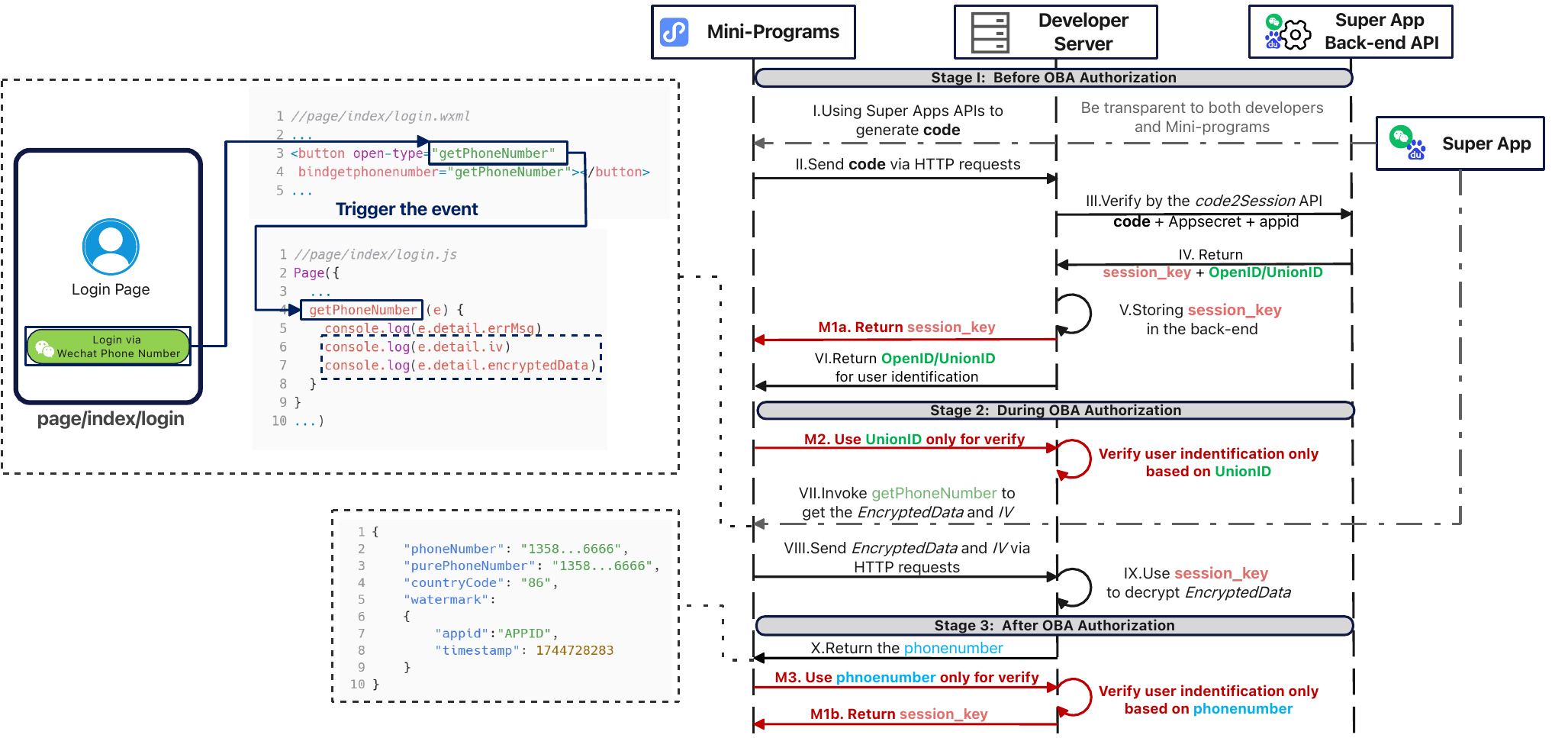}
	\caption{The OBA process in WeChat mini-programs. Other mini-programs follow similar steps. }
 \label{fig:mp_oauth}
\end{figure*}


\paragraphNew{Mini-program OBA Workflow} In practice, platforms such as WeChat and Baidu adopt OBA as the foundation of their authentication mechanism. Mini-programs execute inside an already authenticated super app account, so the authentication workflow is designed to bind the mini-program identity to the currently logged-in super app user while mediating access to platform-held user data through vendor APIs. These adaptations retain the high-level goal of user authentication while reshaping protocol details and integration workflows to fit the sandboxed mini-program ecosystem.

As illustrated in \Cref{fig:mp_oauth}, we take the WeChat mini-program authentication process to demonstrate the OBA workflow. Other platforms follow a similar procedure. When a user launches a mini-program, the front-end first invokes the super app API to request a short-lived, single-use authentication \texttt{authcode}, which is generated internally by the super app (Step I). 

The \texttt{authcode} is then sent to the developer's back-end server (Step II), which forwards it to the super app's authentication back-end (e.g., \texttt{code2Session}~\cite{code2Session}) along with app credentials (Step III). If the request is valid, the super app back-end returns a set of user credentials, including a user identifier (e.g., \texttt{OpenID} and \texttt{UnionID}) and a cryptographic session key \texttt{session\_key} (Step IV). 

The server then returns a response to the front-end, usually excluding the \texttt{session\_key}, which must remain securely stored on the server (Step V). The user identifier (e.g., phone number) is then returned to the front-end for unique user identification (Step VI).

\paragraphNew{Sensitive Data Access via OBA}
Once the OBA process succeeds, the mini-program and its back-end can request sensitive data from the super app, as illustrated in \Cref{fig:mp_oauth}. In Step VII, user interaction (e.g., button click) triggers sensitive data access upon explicit consent.

Upon receiving consent, the mini-program invokes the super app's APIs, such as \texttt{getPhoneNumber}, to retrieve an encrypted payload. This payload includes the sensitive user data (\texttt{encryptedData}) and an associated initialization vector (\texttt{IV}). This data is generated by the mini-program platform and contains information such as the user's phone number or profile data, encrypted using the \texttt{session\_key} obtained in Step IV.

Once the front-end receives the payload, it forwards the \url{encryptedData} and \texttt{IV} to the developer’s back-end server (Step VIII), which performs the decryption (Step IX). The decrypted result typically includes the user’s phone number, country code, and metadata such as a timestamp. This mechanism allows the mini-program to obtain the requested user data securely without directly accessing system resources.

\paragraphNew{Mini-program OBA vs. Standard OAuth/OIDC}
We use the term OBA following vendor documentation~\cite{wechat-login}, but mini-program OBA serves a different usage scenario from standard OAuth 2.0 (RFC 6749) and OpenID Connect~\cite{oauth-rfc6749,openid-connect-core}. Standard OAuth/OIDC commonly supports cross-site or cross-application authorization through browser-visible protocol states, such as redirects, \url{redirect\_uri}, \texttt{state}, authorization endpoints, and signed identity tokens. Mini-program OBA instead operates inside a super app runtime where the user has already logged into the host super app, and the mini-program identity is expected to remain bound to that same super app account.

Accordingly, the front-end invokes vendor APIs such as \texttt{wx.login} or \texttt{swan.login}; the developer server exchanges the resulting \url{authcode} with the vendor's back-end API such as \texttt{code2Session}; and sensitive data is later delivered as an encrypted payload tied to a symmetric \texttt{session\_key}. In this ecosystem, \texttt{OpenID} and \texttt{UnionID} are raw static identifiers rather than OIDC \texttt{id\_token}s. They are not JWT-signed tokens and do not carry nonce-based OIDC guarantees. OBA therefore relies on correct back-end handling of \texttt{session\_key} and platform-returned identifiers rather than a standard token-signature mechanism. This difference does not imply that one design is inherently better than the other; rather, the two designs support different authentication settings. The misuses studied in this paper arise when developers misunderstand the trust boundaries of the mini-program OBA workflow.

%% file: 3.Motivation.tex
\section{OBA Misuse in Mini-programs}
\label{sec:motivation}

In this section, we present the threat model and a taxonomy of runtime OBA misuses in real-world mini-programs.

\subsection{Threat Model}
We assume both the attacker and the victim access mini-programs via official super apps on common platforms such as iOS, Android, and Windows. The attacker has no access to the victim’s device and does not exploit other vulnerabilities to compromise it. However, acting as an active adversary on their own controlled device, the attacker can bypass TLS protections (e.g., via a man-in-the-middle proxy) to monitor and modify outgoing HTTP(S) request parameters. Additionally, the attacker can obtain mini-programs to unpack and decompile their source code. The attacker can also collect auxiliary metadata and public identifiers to facilitate targeted exploitation. Their primary goal is to exploit OBA flaws for victim impersonation, which may inherently result in the disclosure of sensitive data, such as phone numbers or session credentials. We assume the host super app's sandbox and the underlying OS are trusted, focusing solely on logical flaws within the third-party developer's OBA implementation.


\subsection{OBA Misuse Types}
As discussed in Section~\ref{sec:background}, OBA security hinges on correct developer implementation. Beyond hard-coded credentials~\cite{wxsecguideline}, we find that vulnerabilities frequently emerge from the insecure dynamic processing of OBA data during runtime. This identifies a distinct class of risks stemming from logic flaws in the interaction between the mini-program front-end and the developer's back-end server.

To the best of our knowledge, such dynamic behaviors in mini-programs have not been systematically studied.

We analyze the whole lifecycle of OBA shown in \Cref{fig:mp_oauth}, and identify three types of OBA misuses in mini-programs in different steps of the lifecycle.
\paragraphNew{M1: Credential Exchange Before Authorization}
M1 captures failures in credential isolation during the credential-exchange stage or in the subsequent handling of platform-issued session credentials. In a secure OBA flow, the server must be the sole entity responsible for constructing or decrypting identity-related payloads. However, we identify a critical logic flaw where the back-end improperly delegates these cryptographic privileges to the untrusted front-end. This occurs when the server exposes necessary authentication parameters, such as the \texttt{session\_key}, to the client side during the exchange or in a later response, enabling an attacker to forge identity credentials or platform-encrypted data.

(M1a) \textit{Identity Forgery:} By possessing these parameters, the untrusted client can locally encapsulate any victim's identifier into an encrypted payload. When this forged payload is sent in Step VIII, the server accepts it as a legitimate, platform-verified identity. 
This constitutes a dynamic authentication bypass where the attacker dictates the user identity, rather than the platform.

(M1b) \textit{Post-authentication Data Forgery:} If the \texttt{session\_key} is exposed after the initial login, attackers cannot forge the immediate authentication request, but OBA credential isolation is still broken. Attackers can decrypt, modify, and re-encrypt subsequent platform payloads. For example, by altering the \texttt{encryptedData} from APIs like \texttt{getUserInfo}, an attacker can forge profile attributes to bypass certain account restrictions. Though weaker than M1a, M1b captures the bypass of subsequent authorization checks. This split refines prior leakage findings~\cite{DBLP:conf/ndss/ShiYZ0Y0025,chen2025whiskey} by distinguishing attack timing within the OBA lifecycle.

\paragraphNew{M2: Identity Assertion During Authorization} M2 occurs during the identity-assertion stage, where the back-end should accept an identity only when it is bound to a fresh OBA exchange. The UnionID is a platform-assigned identifier unique to a user across applications of the same developer~\cite{unonid-wechat,unonid-baidu}. Secure OBA implementations use UnionID only for data mapping, yet we find many mini-programs treat it as a standalone authentication credential.

This flaw ignores the requirement for dynamic cryptographic verification. Because the UnionID is static, an attacker can obtain it from a less secure application under the same vendor and reuse it for impersonation. By submitting the UnionID directly to the server, the attacker bypasses the OBA flow entirely. This transforms a public identifier into a permanent secret, failing to ensure that the request originates from the legitimate owner.

\paragraphNew{M3: Channel Request After Authorization} M3 occurs after OBA has retrieved sensitive data for the mini-program, when that data is propagated to later application logic. In some implementations, developers use OBA solely to fetch sensitive data, such as phone numbers. However, this data is subsequently transmitted in plaintext without cryptographic protection. By modifying these plaintext identifiers in transit, an attacker can bypass authentication and impersonate arbitrary users, as the back-end lacks a verification mechanism to ensure the data's authenticity.

\begin{figure}[t]
	\centering
	\includegraphics[width=1\linewidth]{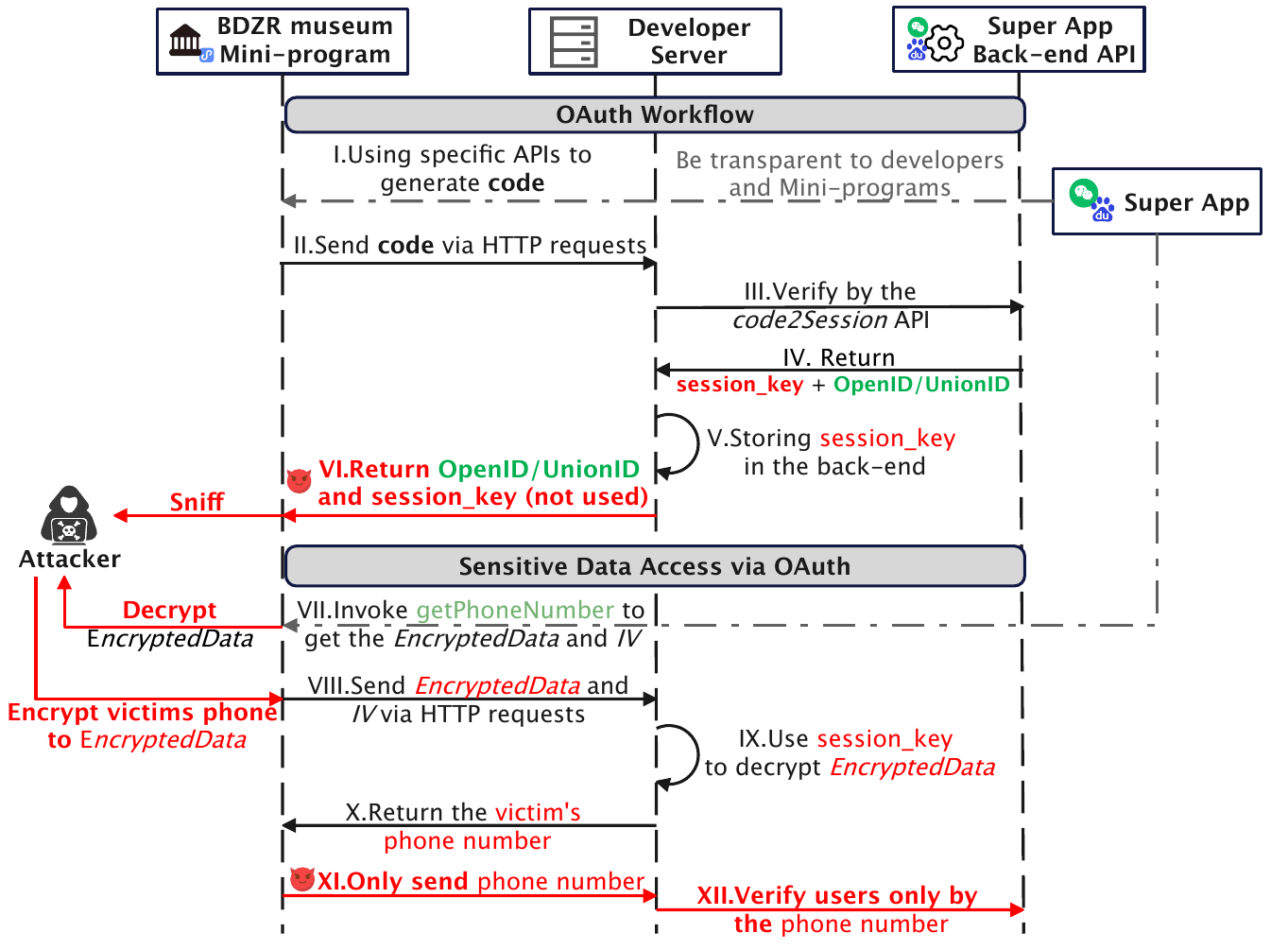}
	\caption{Misuses and attack process of the BDZR Museum mini-program.}
 \label{fig:motivation_bdzr}
\end{figure}

\begin{figure}[tbh]
	\centering
 \includegraphics[width=1\linewidth]{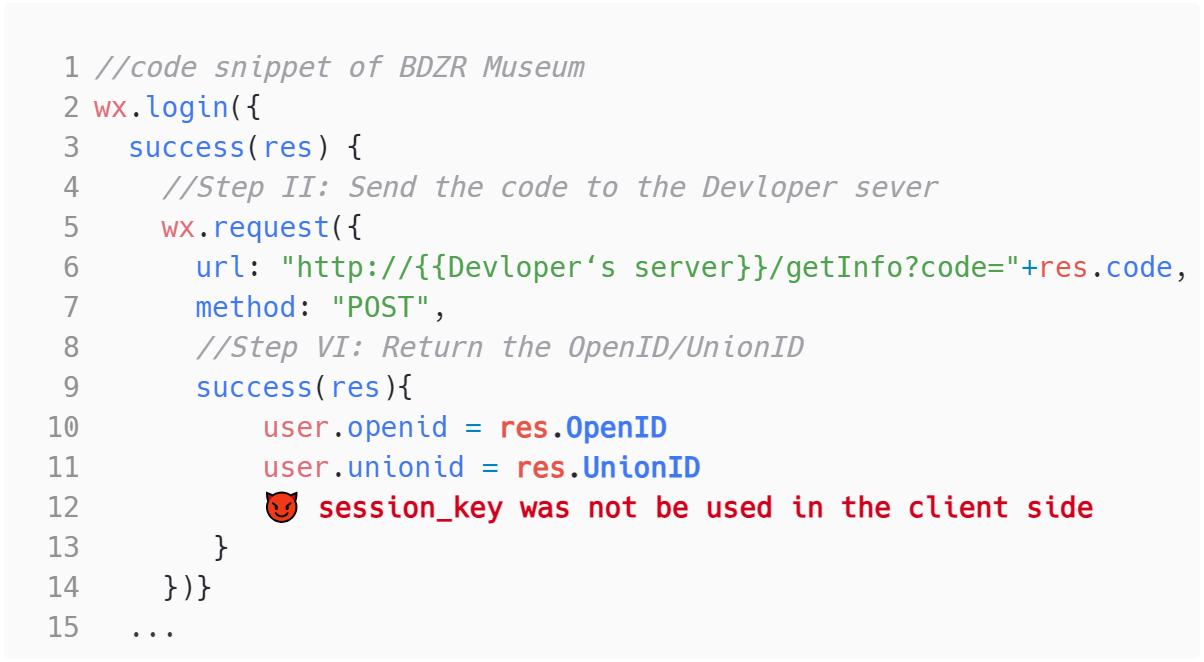}
	\caption{The code snippet of the OBA process from the BDZR Museum mini-program.}
 \label{fig:motivation_bdzr_code}
\end{figure}

\subsection{Motivating Example}
\label{subsec:motiexample}

We present a real-world case study involving the \textit{BDZR Museum}\footnote{The vulnerability has been fixed at the time of paper submission.} mini-program to illustrate vulnerabilities arising from two types of OBA misuse.

The first five steps are the same as we discussed in \Cref{sec:background}.
The first misuse comes in Step VI: in addition to user IDs (\texttt{OpenID} and \texttt{UnionID}), the developer's server also returns the session key to the mini-program front-end (\textbf{M1}), as shown in \Cref{fig:motivation_bdzr}.
When an attacker is running this mini-program on their device, they can obtain the session key by sniffing.
Later in Step VII, the attacker can decrypt the \texttt{EncryptedData} with the obtained session key and IV to get the plaintext that contains the attacker's ID (e.g., the phone number).
Since the attacker has the session key and IV, they can generate a new EncryptedData object by replacing the attacker's ID with the victim’s and resend it to the server (Step VIII).
As a result, the attacker can successfully log in to the mini-program using the victim's ID.

The second misuse occurs in Step XI, where the user's ID obtained from the previous step is sent to the developer's server in plaintext (\textbf{M3}).
This allows the attacker to directly replace her ID with the victim's in the request body, completely bypassing encryption-based defenses.




%% file: 4.Challenges.tex
\section{Challenges and Solutions}
\label{sec:challenges}

Existing static approaches fail to detect runtime OBA misuses, even when bypassing platform-level checks~\cite{sessionkeyclasswx}. As shown in \Cref{fig:motivation_bdzr_code}, sensitive parameters like \texttt{session\_key} can be injected into network traffic without appearing in the front-end source code. This runtime disparity, where critical vulnerabilities leave no static footprint, necessitates dynamic analysis but introduces non-trivial challenges. We detail these challenges and our solutions below.

\paragraphNew{Limitations of Current OBA-related Work}
Mini-programs differ fundamentally from traditional web and mobile applications. They operate within the sandbox of a host super app, meaning authentication is inherently tied to the active super app account. Consequently, a user's identity is determined by the super app login rather than an external website. While mini-program providers maintain independent back-end user states, OBA serves to bridge these states with the authenticated super app account. Unlike web-based OAuth, mini-program OBA does not involve parameters such as \texttt{redirect\_url} or \texttt{state}. Furthermore, since all communication occurs within the super app environment, it does not rely on inter-app broadcasts or intents common in mobile SDK-based OBA. These characteristics define mini-program OBA as a distinct problem space where existing research in Web and mobile OAuth security (e.g., studies focused on open redirect or inter-app hijacking) cannot be directly applied, as the underlying attack surfaces and protocols are unique to the super app ecosystem.

One line of work that is close to this topic is web SSO testing.
In SSO testing, tools can reason about authentication by observing browser-visible redirects, URLs, DOM elements, cookies, and OAuth parameters.
However, in mini-program OBA, the security-critical transitions are mediated by super-app JS Bridge calls, native authorization dialogs, and vendor APIs.
These states are not indicated in DOM elements or redirect states, preventing SSO testing techniques to apply.


Additionally, existing approaches for analyzing mini-programs~\cite{wang2023taintmini,zhang2023don} face two main limitations regarding OBA misuse detection. First, static approaches struggle to analyze obfuscated mini-programs, which are prevalent in real-world deployments. Second, static methods cannot interact with mini-programs at runtime. They fail to capture dynamic behaviors not reflected in the source code, such as the misuses discussed in \Cref{subsec:motiexample}. These limitations highlight the need for a dynamic, interaction-driven approach that can automatically execute the mini-program and monitor its behaviors.

\paragraphNew{Challenges and Solutions} To evaluate the impact of discovered OBA runtime misuses, dynamic analysis is necessary. However, detecting OBA misuse at scale presents two key technical challenges.

The first challenge stems from code obfuscation. To automatically detect OBA misuse, the framework must first pinpoint the OBA login page and trigger the authentication process. However, \textbf{(C1) how to automatically identify the OBA login page in obfuscated mini-programs} is non-trivial. Modern obfuscation techniques obscure page routing, event handlers, and component metadata. This prevents static analysis from mapping UI elements to OBA-related logic or determining which page serves as the entry point for authentication. Furthermore, obfuscated code often employs dynamic path generation and encrypted resource loading, which further hinders any attempt to identify login-related components through static inspection. Consequently, identifying the login interface requires dynamic exploration rather than static analysis.

{\bf Solution:} To address this challenge, we first extract the complete set of registered pages from the configuration file of a mini-program. Then, we automatically enumerate and explore each page using UI automation and OCR-based keyword recognition to identify the OBA login page as the entry point for further automatic analysis. We design a traversal algorithm that starts from the index page and keyword-matched routes. This algorithm explores clickable UI components in a DFS-like manner and revisits pages when new elements appear after navigation. Details are provided in \Cref{subsec:loginpageidentifier}.

The second challenge originates from automation. Mini-programs run inside super apps like WeChat and Baidu. Unlike mobile applications that allow direct interaction via standard interfaces such as Android Intents, mini-programs are sandboxed within the host application. This architecture lacks external interfaces for direct manipulation. Consequently, \textbf{(C2) how to automatically interact with mini-programs} remains a significant hurdle. Traditional dynamic analysis techniques are restricted because OBA flows in mini-programs depend on tightly coupled platform APIs and real-time user interactions that occur within a closed environment.

{\bf Solution:} To address this challenge, we reverse-engineer the control paths within super apps and design an external automation framework. This framework enables direct interaction with mini-programs from outside the super app to facilitate automatic execution. We implement an Xposed plugin that hooks into the super app process to craft broadcast Intents or deeplinks containing the target \texttt{AppID} and page \texttt{Path}. These are then invoked by the super app to launch the target mini-program. Our framework can launch a mini-program, navigate to specific pages, and interact with front-end UI elements. Moreover, it requires neither source code instrumentation nor platform SDK support, as detailed in \Cref{subsec:dynamicanalyzer}.

%% file: 5.Design.tex
\section{Design of \sysname} 
\label{sec:design}

\begin{figure*}[t]
	\centering
	\includegraphics[width=1\textwidth]{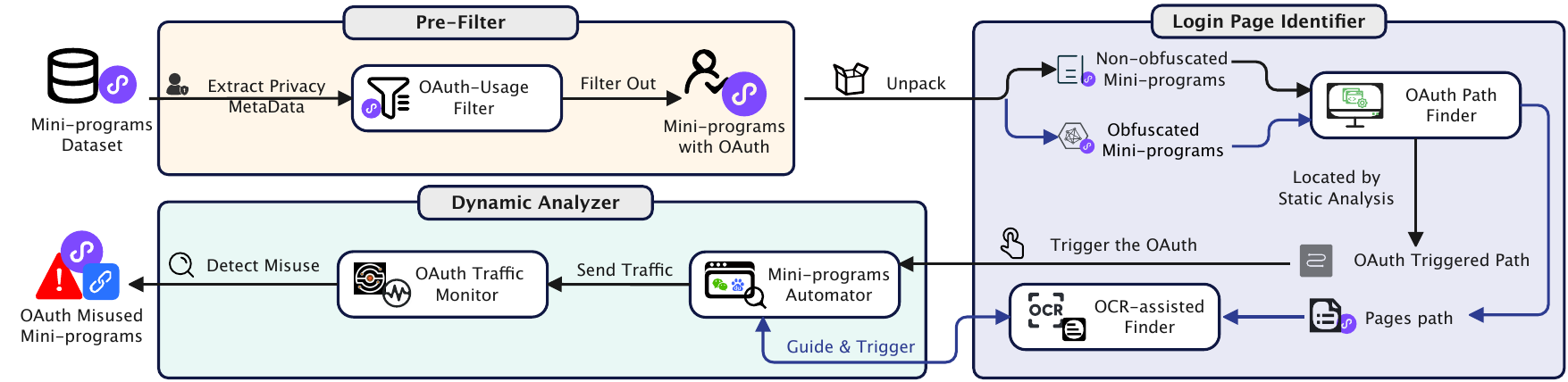}
	\caption{The workflow of \sysname.} 
	\label{fig:design_workflow}
\end{figure*}

To address the limitations in the existing work and detect OBA misuse issues at a large scale, we designed \sysname, the first dynamic analysis framework for automatic analysis of mini-programs.
As illustrated in Figure~\ref{fig:design_workflow}, \sysname consists of three main components: \textit{Pre-filter}, \textit{Login Page Identifier}, and \textit{Dynamic Analyzer}.
As the first step, the Pre-filter filters out mini-programs without OBA from all the crawled mini-programs.
Then, the Login Page Identifier analyzes the mini-program either statically or dynamically to pinpoint the OBA login page in the mini-program.
Lastly, the Dynamic Analyzer takes the login page as the entry point and automatically executes the OBA workflow to capture and analyze the traffic.
\sysname currently supports both WeChat and Baidu mini-programs, covering the two most widely used platforms.

\begin{algorithm}[t]
\small

\caption{Finding OBA event trigger elements in the render layer.}
\begin{algorithmic}[1]
    \Function {findOBAEvent}{staticFile, jsFile, EventFunction}
        \State newSourceCode $\gets$ $\emptyset$
        \State newSourceCode += (convertoHTML(staticFile), jsFile)
        \State newAST $\gets$ $\emptyset$
        \State newAST += buildAST(newSourceCode)
        \State SinkNode = EventFunction
        \State SourceNode = reverseTaint(newAST, SinkNode)
        \State TriggerElement = SourceNode
        \State \Return TriggerElement
    \EndFunction
\end{algorithmic}
\label{alg:algo_new_ast}
\end{algorithm}

\subsection{Pre-filter}
\label{subsec:prefilter}

Due to regulatory requirements, mini-programs must publicly disclose their intent to collect sensitive user information such as phone numbers, dates of birth, and gender. In the mini-program ecosystem, OBA serves as the exclusive mechanism for accessing such sensitive user data. Consequently, the official guidelines~\cite{user-privacy-mp-wechat} mandate that developers must provide a corresponding privacy declaration specifying the categories and purposes of data collection before utilizing OBA APIs. This declaration specifies the categories of user information to be collected and the purposes of collection. Consequently, we use the privacy declaration as a reliable indicator of OBA usage. We note that the set of mini-programs with such declarations is a superset of those actually utilizing OBA, as some developers may preemptively publish declarations to ensure policy compliance even if the OBA flow is not yet active. By crawling these standardized privacy pages via \texttt{AppID} and applying pattern matching, we can effectively identify potential OBA-enabled mini-programs while filtering out those that do not handle sensitive data.


Each program’s \texttt{AppID} links to a structured URL hosting its declaration, which follows a standardized format. For example, the disclosure for \texttt{getPhoneNumber} typically states that the developer will collect the user’s phone number with explicit consent. Leveraging this consistency, we crawl the privacy pages via \texttt{AppID} and apply pattern matching to identify potential mini-programs using OBA while filtering out those that do not.

\begin{algorithm}[htbp]
\small
\caption{Identifying OBA entry points via traversal.}
\begin{algorithmic}[1]
    \Function{FindOBAEntry}{$ConfigFile$}
        \State $Routes \gets Parse(ConfigFile)$
        \State $Candidates \gets \{Homepage\} \cup KeywordFilter(Routes)$
        \State $Visited \gets \emptyset$
        \While{$Candidates \neq \emptyset$}
            \State $Page \gets Pop(Candidates)$
            \If{$Page \in Visited$} 
                \State \textbf{continue} 
            \EndIf
            \State $Visited \gets Visited \cup \{Page\}$
            \ForAll{$Element \in EnumerateUI(Page)$}
                \If{$IsClickable(Element)$}
                    \State $Result \gets Click(Element)$
                    \If{$Result = OBAPopup$}
                        \State \Return $Page$  \Comment{OBA entry identified}
                    \ElsIf{$Result = NewPage$}
                        \State $Candidates \gets Candidates \cup \{NewPage\}$
                    \EndIf
                \EndIf
            \EndFor
        \EndWhile
        \State \Return None  \Comment{No OBA entry found}
    \EndFunction
\end{algorithmic}
\label{alg:oauthfinder}
\end{algorithm}

\subsection{Login Page Identifier}
\label{subsec:loginpageidentifier}

After identifying mini-programs that use OBA, \sysname unpacks the target applications. If the program is not obfuscated, \sysname statically analyzes the source code to pinpoint the OBA login page. In cases where the mini-program is obfuscated, \sysname employs an OCR-assisted finder to systematically enumerate and explore each page within the mini-program to identify the OBA login entry.

\paragraphNew{Non-obfuscated Mini-program} To identify OBA-related interactions in non-obfuscated mini-programs, we perform static analysis across the logic layer (e.g., \texttt{.js} files) and the render layer (e.g., \texttt{.wxml} or \texttt{.swan} files). Since these render files closely resemble HTML, we adopt a strategy similar to that of web application analysis by linking UI event bindings to their corresponding logic handlers.

The analysis starts by locating platform-specific OBA APIs (i.e., \texttt{wx.login}, \texttt{swan.login}, \texttt{getPhoneNumber}, and \texttt{getUserInfo}) in the logic layer. We then trace their handler functions and match them to UI elements via event attributes such as \texttt{bindtap} and \texttt{bindgetphonenumber}. This allows us to pinpoint which components (e.g., \texttt{<button>} tags) trigger the OBA flow. \Cref{alg:algo_new_ast} illustrates how function references are traced and resolved to event bindings.
The discovered UI element and its page path are recorded as the OBA entry point for downstream analysis.

\paragraphNew{Obfuscated Mini-program} If a mini-program's source code is obfuscated, the above-discussed source-level analysis is no longer feasible. 
However, platform conventions still mandate that all page routes be explicitly declared in a configuration file that is neither obfuscated nor encrypted, such as the \texttt{app.json} file in WeChat.
This file serves as a high-level blueprint to traverse all pages in the mini-program. Our key observation to identify the trigger of the OBA process is that once it is correctly triggered, the super app displays a fixed-style pop-up dialog, such as an authorization window asking the user to ``Login'' or ``Authorize''. As illustrated in Figure 6, these platform-controlled dialogs exhibit consistent UI patterns across different mini-programs, which provides a reliable and observable signal for identifying the OBA entry point during our automated exploration.

To automate the traversal process, a naive approach would be to explore all declared pages and elements exhaustively. 
However, this approach is prohibitively time-consuming, making large-scale analysis infeasible.
Instead, \sysname optimizes the traversal process as follows:

\begin{enumerate}
    \item \textit{Step I}: We begin with the homepage and all keyword-matched routes (e.g., \texttt{login}, \texttt{auth}), where UI elements are enumerated and clickable candidates are tested.
    \item \textit{Step II}: We use a depth-first search strategy to drive the traversal process. If a click successfully triggers the OBA popup, the process terminates; otherwise, if it navigates to a new page, we recursively explore that page until the OBA popup appears or no further elements can be clicked. 
    \item \textit{Step III}: If the OBA pop-up still cannot be triggered after step (2), we re-enter the traversal process on previously visited pages. Since some clickable elements only appear after specific navigation paths, this ensures that secondary paths, which may not be visible during the first visit, are also explored.
    
\end{enumerate}

This recursive, coverage-oriented traversal allows \sysname to efficiently identify OBA entry points even under heavy obfuscation.
The process is formalized in Algorithm~\ref{alg:oauthfinder}.

\subsection{Dynamic Analyzer}
\label{subsec:dynamicanalyzer}

Once the entry point of a mini-program is identified, \sysname automates the end-to-end process of interacting with mini-programs and capturing OBA behavior.

\subsubsection{Mini-program Automator} 
To dynamically execute OBA flows at scale, we need to first automatically navigate to the login page within a mini-program. However, unlike standalone mobile apps, mini-programs are hosted within super apps, which do not expose public APIs for launching arbitrary mini-programs from outside the super app. This introduces unique challenges for automating page-level navigation.

To address this issue, we reverse-engineer the super apps to uncover how mini-programs are loaded and how parameters such as \texttt{AppID} and page \texttt{Path} are propagated at runtime. On WeChat, we find that the official API~\cite{wx-navigateToMiniProgram} is invoked via external \texttt{Intent} messages on Android. Based on this finding, we implement an Xposed plugin that injects crafted Intents with the target \texttt{AppID} and page \texttt{Path} into the WeChat process. On Baidu, Alipay, and TikTok, we discover that mini-programs can be launched via standardized deep link formats~\cite{w3c-schema}, which encode the target parameters into a URI.

Based on these findings, we implement an automation module that simulates native navigation behavior. When launching a mini-program, our framework either hooks into internal APIs or constructs valid deep links to open specific pages. This approach is generic and extensible; while the reverse-engineering of launch mechanisms requires a one-time effort per platform, it enables the automated analysis of all mini-programs within that ecosystem without developer-side permissions or manual intervention.

Upon reaching the designated OBA login page, \sysname initiates a structured interaction process to start the OBA workflow.
This process is designed to be platform-agnostic by properly handling the following three scenarios:

\begin{figure}[t]
	\centering
	\includegraphics[width=1\linewidth]{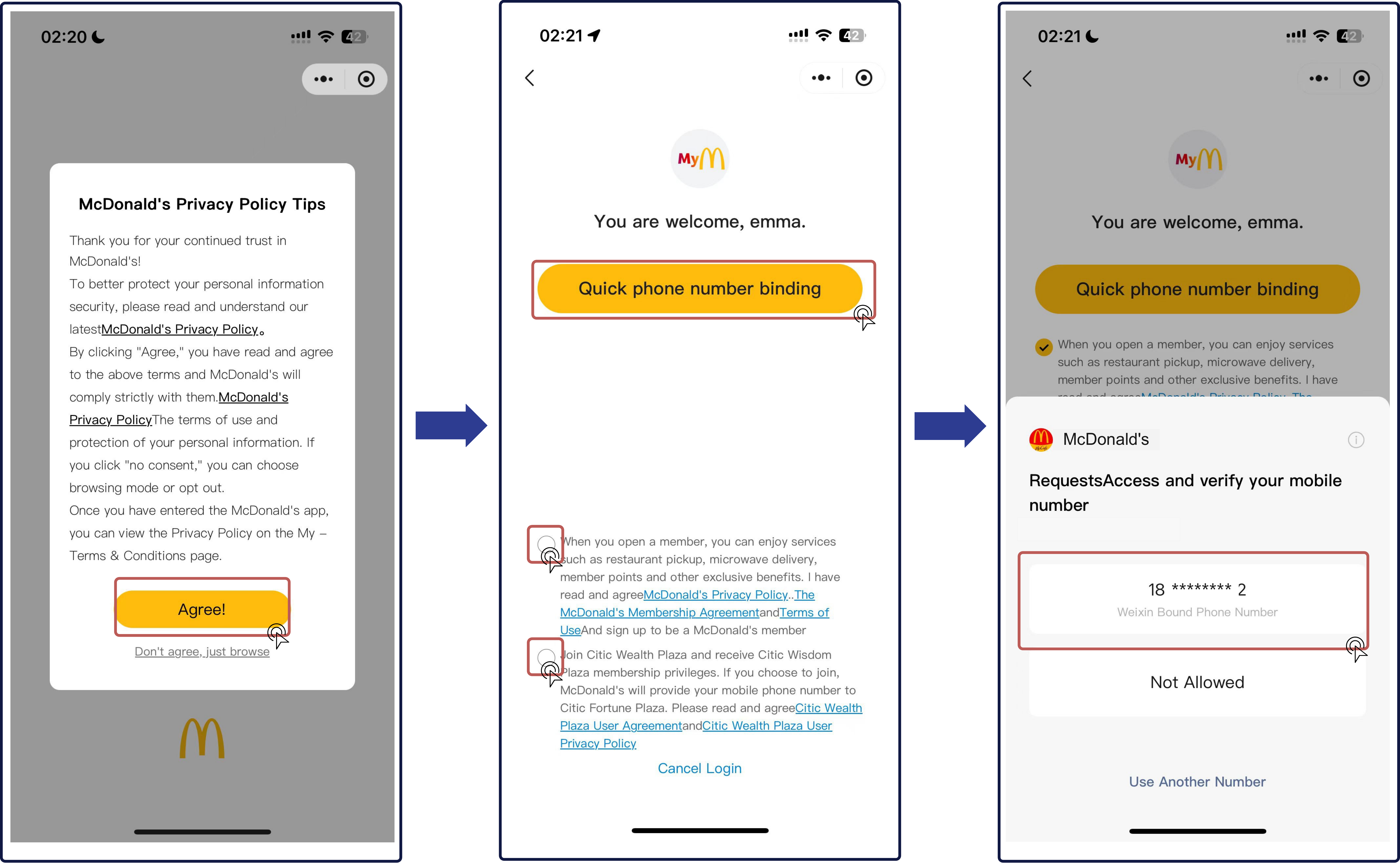}
	\caption{An example of how to trigger the OBA process in WeChat mini-programs.} 
	\label{fig:grant_privacy}
\end{figure}



\noindent{\bf Platform Authorization Prompts.}
  When a user clicks a login button in a mini-program, the super app often displays a built-in pop-up requesting permission to access sensitive data, such as the phone number, as seen in the \texttt{Agree!} button of \Cref{fig:grant_privacy}. The super app itself manages these pop-ups and cannot be triggered by clicking directly on the mini-program buttons. To handle this, our system detects when such a pop-up appears on screen and simulates the user's click to approve the request, allowing the login process to continue without manual assistance.

\noindent{\bf Privacy Agreements Pop-up.}  
    Before login can proceed, many mini-programs present developer-defined privacy terms.
    These typically follow two interface styles: checkbox-based agreements, such as the two checkboxes in \Cref{fig:grant_privacy}, and modal pop-up confirmations like the phone number request pop-up toast in \Cref{fig:grant_privacy}.
    \sysname includes a generic detection routine that searches for checkable elements and monitors the dialog window for clickable components.
    It automatically acknowledges the agreement by interacting with the relevant UI components.

\noindent{\bf OBA Flow Execution.}  
    Once the login is triggered, the system proceeds through the platform-provided OBA interface.
    While the specific user data being requested may differ across mini-programs (e.g., phone number and/or profile information), the interaction model is uniform, enabling robust automation at this stage.

By addressing these three scenarios, \sysname automatically executes the OBA workflow while capturing and analyzing the resulting network traffic to identify dynamic security behaviors.

\subsubsection{OBA Traffic Monitor} 
To identify potential misuse of OBA mechanisms, \sysname monitors all network traffic during the dynamic execution of mini-programs. During monitoring, once certain parameters are detected in the traffic, \sysname identifies this mini-program as vulnerable.

\sysname keeps monitoring the parameters listed in \Cref{tab:oauth-rules}.
Instead of relying on simple string matching, \sysname adopts a \textit{context-based template detection} approach. The key idea is to leverage both the syntactic patterns of parameter values and their structural roles in request/response payloads, thereby inferring security-critical fields even when names are randomized. Specifically, \sysname builds detection templates consisting of two dimensions:  

$\bullet$ \textit{Contextual Format Templates.}  
    Each sensitive OBA field (e.g., \url{session\_key}, \url{access\_token}, \url{UnionID}) exhibits a highly constrained structural pattern, such as fixed lengths, base64-like encoding, or numerical ranges. \sysname encodes these constraints into reusable format templates, which are applied not in isolation but in the context of where the parameter appears (e.g., embedded in JSON payloads, HTTP headers, or form-encoded bodies).

$\bullet$ \textit{Semantic Name Templates.}  
    In addition, \sysname maintains a set of evolving name templates that capture the common variants that developers use to denote the same field. For instance, \texttt{session\_key} may occur as \texttt{skey}, \texttt{sessionKey}, or other fuzzed forms. These templates are continuously refined through prior detections and combined with fuzzy matching, allowing the system to accommodate developer-specific naming conventions.


By jointly considering the contextual placement of parameters and their template-constrained value semantics, this hybrid detection mechanism achieves higher robustness than pure regex-based rules. Concretely, \sysname determines that a mini-program is vulnerable to specific misuses under the following conditions:

$\bullet$ \textit{M1: Client-side Identity Forgery.} 
To detect M1 vulnerabilities, \sysname monitors whether parameters required for identity construction, such as \texttt{session\_key} and \texttt{IV}, are exposed to the client-side. The detection logic distinguishes between the two sub-categories based on the timing of exposure relative to the authentication request:

\begin{itemize}
    \item \textit{M1a (Identity Forgery):} \sysname flags this when the required credentials are exposed \textit{before} the request containing \texttt{encryptedData} is sent (e.g., in Step VI). In this case, \sysname automatically attempts to decrypt the \texttt{encryptedData} using the intercepted credentials. A successful decryption confirms that the front-end has the capacity to construct identity payloads, enabling active identity forgery.
    \item \textit{M1b (Post-authentication Data Forgery):} If these credentials are only detected in network traffic \textit{after} the ciphertext has been transmitted, \sysname categorizes the case as M1b. While the attacker may lack the immediate opportunity to forge the current authentication request, this exposure breaks credential isolation and may allow subsequent platform-encrypted payloads to be decrypted, modified, or re-encrypted before being accepted by the back-end.
\end{itemize}

$\bullet$ \textit{M2 Detection.}   
After the OBA flow completes, \sysname flags M2 when the observed traffic contains requests carrying a parameter that matches the \texttt{UnionID} template, but no accompanying \texttt{OpenID}. 
This condition indicates that the mini-program relies solely on \texttt{UnionID} as the authentication credential, which is a misuse of OBA and enables trivial impersonation attacks across applications.

$\bullet$ \textit{M3 Detection.}   
After the OBA flow completes, \sysname flags M3 if traffic contains requests where sensitive data (e.g., phone numbers) is transmitted in plaintext without additional encryption. 
In practice, this appears as phone numbers directly exposed in the request body, without server-side binding to platform identifiers such as \texttt{OpenID} or \texttt{UnionID}.

\begin{table}[t]
\centering
\caption{Template-based detection rules.}
\label{tab:oauth-rules}
\small
\begin{tabular}{lccc}
\toprule
\textbf{Parameter} & \textbf{Pattern} & \textbf{Name} & \textbf{Scope} \\
\midrule
Phone \#        & 1[3-9]\d\{9\}               & No  & Req \\
session\_key    & [A-Za-z0-9+/]\{22\}==       & Yes & All \\
IV              & [A-Za-z0-9+/]\{22\}==       & Yes & All \\
EncryptData     & [A-Za-z0-9+/]\{214\}==      & Yes & All \\
access\_token   & [A-Za-z0-9./=\-\_]\{200,300\}& Yes & All \\
UnionID         & .{28}                     & Yes & All \\
\bottomrule
\end{tabular}
\end{table}

%% file: 7.Implementation.tex
\section{Implementation of \sysname} 
\label{sec:implementation}

\sysname supports the analysis of both WeChat and Baidu mini-programs.
Our Pre-filter uses a Python crawler to fetch declaration pages via \texttt{AppID} and applies regex-based OBA keyword matching to filter out mini-programs without OBA.
The Login Page Identifier is implemented based on CodeQL~\cite{codeql-js} to statically
identify the login pages and the UI elements that trigger OBA. 

To automatically start the mini-program, we implement an Xposed plugin for WeChat that hijacks internal calls to \url{wx.navigateToMiniProgram}~\cite{wx-navigateToMiniProgram} by injecting \texttt{AppID} and \texttt{Path} via intent. 
Xposed~\cite{xposedbridge} serves solely as a WeChat launch adapter to trigger native navigation paths. It remains completely decoupled from OBA credentials, code modification, and \sysname's core detection logic. 
For Baidu, \sysname uses Appium~\cite{appium} to launch mini-programs via \texttt{baiduboxapp://} deep links~\cite{baidu-schema}. 
Because \sysname isolates this launch mechanism, porting to other super apps or adapting to vendor updates only requires modifying the platform-specific adapter.

To execute the OBA workflow, \sysname uses Appium to interact with UI elements and an OCR-powered Appium plugin~\cite{appium-ocr-plugin} to detect keywords on the UI. 
For platform-controlled one-click authorization or privacy-consent dialogs, \sysname identifies common consent actions and triggers them using our test account; mini-programs requiring manual identity checks or account-specific inputs are marked as inaccessible for end-to-end analysis.
\sysname employs \texttt{Mitmproxy}~\cite{mitmproxy} to capture and analyze HTTPS traffic during OBA.

%% file: 6.Evaluation.tex
\section{Evaluation}
\label{sec:evaluation}

In this section, we evaluate \sysname’s effectiveness and coverage through large-scale measurements and case studies.


\subsection{Dataset \& Experiment Setup}

\paragraphNew{Dataset} 
Our study focuses on WeChat and Baidu because WeChat is the biggest and most popular mini-program platform. At the same time, Baidu hosts more diverse mini-programs~\cite{baidu-cate} and supports an open-source alliance~\cite{baidu-opensource}, making it more representative and increasing the potential impact of misuse. 
We crawled WeChat mini-programs by simulating user interactions on the Windows client, extracting mini-program packages and \texttt{AppIDs} from the local \texttt{user\_file/Applet/\{AppID\}} directory. For Baidu, we automated in-app searches on an Android phone and retrieved mini-program files and \texttt{AppIDs} from \texttt{aiapps\_folder/\{AppID\}} in internal storage.
Our dataset comprises a total of 44,273 WeChat mini-programs and 2,721 Baidu mini-programs, occupying approximately 129.5 GB of disk space.
We note that increasingly strict anti-crawling measures (such as rate limiting, JS-based verification, and dynamic token challenges) deployed by both Tencent and Baidu make acquiring large-scale mini-program datasets significantly more difficult than in earlier studies~\cite{zhang2021measurement}.


Nevertheless, \sysname benefits from the runtime nature of dynamic analysis. Due to the hot-update mechanism of mini-programs, our system constantly interacts with the latest version available on vendor platforms. This ensures that our detection results reflect the most current application logic and security posture, which is not achievable through prior static analysis efforts. In contrast, static approaches typically analyze datasets that only capture the state of mini-programs at the time they were crawled.
This enables \sysname to capture recent security updates or regressions, offering greater practical utility than static snapshots.

\paragraphNew{Experiment Setup}
We deployed and ran the static pre-filter of \sysname on a Windows 11 PC equipped with an Intel i7-13700 CPU and 64 GB of RAM.
We utilized 10 parallel processes to analyze all unpacked mini-programs in our dataset efficiently.
For the dynamic analyzer of \sysname, we performed experiments on a rooted Pixel 4 device running Android 13, which allowed us to connect to the WeChat and Baidu app environments and perform the low-level instrumentation required by \sysname.

\begin{table}[htbp]
\centering
\caption{OBA misuse cases detected by \sysname across WeChat and Baidu mini-programs. Affected mini-programs may match multiple categories.}
\label{tab:overview_res}
\begin{tabular}{lrrrrrr}
\toprule
\textbf{Vendor} & \textbf{Mini-program Type (\#)} & \textbf{M1a} & \textbf{M1b} & \textbf{M2} & \textbf{M3} \\
\midrule
\multirow{2}{*}{WeChat} 
    & Non-obfuscated (5,342) & 427 & 719 & 157 & 219 \\
    & Obfuscated (4,817)     & 39  & 16    & 100 & 63  \\
\cmidrule(lr){1-6}
\textbf{Total} 
    & \textbf{10,159} & \textbf{466} & \textbf{735} & \textbf{257} & \textbf{282} \\
\midrule
\multirow{2}{*}{Baidu}  
    & Non-obfuscated (315)   & 14  & 18    & 2   & 30  \\
    & Obfuscated (326)       & 4   & 15    & 0   & 11  \\
\cmidrule(lr){1-6}
\textbf{Total}  
    & \textbf{641} & \textbf{18} & \textbf{33} & \textbf{2} & \textbf{41} \\
\bottomrule
\end{tabular}
\end{table}

\subsection{Result Overview}
\label{subsec:resultoverview}

Table~\ref{tab:overview_res} summarizes the misuse detection results. \sysname identified 22.9\% (10,159/44,273) of WeChat mini-programs and 23.6\% (641/2,721) of Baidu mini-programs as using OBA. Among them, 13.8\% (1,397/10,159) of WeChat and 45.4\% (291/641) of Baidu mini-programs exhibited at least one type of OBA misuse. Furthermore, 5.7\% (583/10,159) of WeChat and 5.6\% (36/641) of Baidu mini-programs suffered from two or more misuse types.

Additionally, among mini-programs with OBA, 5,342 out of 10,159 (52.6\%) WeChat and 315 out of 641 (49.1\%) Baidu mini-programs are non-obfuscated. The remaining 4,817 WeChat and 326 Baidu mini-programs are heavily obfuscated; this high prevalence of obfuscation highlights the inherent limitations of static analysis and underscores the necessity of \sysname's comprehensive dynamic approach for robust mini-program analysis, including login page identification and runtime logic extraction.

The most prevalent misuse type is \textbf{M1}, which involves the exposure of parameters intended for server-side use to the untrusted client.  In detail, 466/10,159 (4.59\%) of these WeChat and 18/641 (2.81\%) of these Baidu mini-programs are further classified as \textbf{M1a}, as they also expose \texttt{encryptedData} and \texttt{IV} before the authentication request. This sequence enables attackers to perform client-side identity construction and successfully impersonate victim users.
Moreover, 735/10,159 (7.22\%) WeChat and 33/641 (5.15\%) Baidu mini-programs suffer from \textbf{M1b}, where credentials such as the \texttt{session\_key} are exposed to the front-end after the current authentication exchange. These cases usually do not enable immediate login identity forgery as M1a does, but they still break OBA credential isolation and may enable forgery of subsequent platform-encrypted payloads.

In contrast, \textbf{M2} and \textbf{M3} are less frequent as they typically stem from deliberate but insecure design choices. These include relying solely on static identifiers like \texttt{UnionID} for authentication (\textbf{M2}) or transmitting sensitive data in plaintext (\textbf{M3}). We identified such patterns in 257 WeChat and 2 Baidu mini-programs for \textbf{M2}, and 282 WeChat and 41 Baidu mini-programs for \textbf{M3}.


\paragraphNew{Performance and Scalability} We measure the end-to-end processing cost of \sysname after static pre-filtering. On average, \sysname takes approximately two minutes per mini-program, including page launch, OBA-entry triggering, one-click consent handling, traffic collection, and misuse detection. Most of this time is spent waiting for UI rendering, network responses, and platform-controlled authorization dialogs; traffic parsing and rule matching add little overhead. The static pre-filter is lightweight and parallelizable with 10 worker processes in our setup, while the dynamic analyzer uses timeouts and marks mini-programs requiring CAPTCHAs, real-name verification, payment, or other manual interactions as inaccessible. This makes \sysname suitable for offline auditing and vulnerability triage, though not for real-time platform enforcement.

\subsection{Flawed OBA Design from Vendor} 
While the majority of OBA misuse cases originate from insecure developer practices, we also identify flawed designs at the platform level that introduce serious cryptographic weaknesses.
We discover that Baidu mini-programs use a 128-bit \texttt{session\_key} derived via AES-CBC with PKCS7 padding during OBA authentication. More critically, the IV is not randomly generated but reuses the first 42 bits of the \texttt{session\_key} instead. Since both the \texttt{encryptedData} and IV are transmitted over the network, an attacker can mount an offline brute-force attack to recover the full \texttt{session\_key}, even if the developer follows all recommended practices.

To validate this risk, we used an open-source \textit{session\_key} tool \texttt{aes-brute-force}~\cite{aes-brute-force} to accelerate the brute-force process using Intel AES-NI instructions~\cite{aes-ni}. In a controlled experiment, we launched a proof-of-concept attack on a test server equipped with an NVIDIA Tesla-series GPU (16 GB), 8-core vCPU, and 32 GB of RAM. Based on our experimental setup, the computational cost of a successful brute-force session\_key recovery is approximately \$2.74 (derived from the 20-minute execution time on a GPU-accelerated instance ), making the attack both feasible and economically practical for adversaries.
We have responsibly reported our findings to Baidu, which has confirmed the issue and is currently working on a fix.

We also note that WeChat uses a similar cryptographic misstep: its IV likewise reuses part of \texttt{session\_key}. Due to the use of a 256-bit session key, we were unable to complete a brute-force attack within the validity window under current hardware constraints.
Nevertheless, this design leaks part of the session\_key and thus significantly undermines the confidentiality guarantees of encryption.
We also reported this issue to Tencent, and they have confirmed and fixed it.

\subsection{Accuracy of \sysname Detection} 
To evaluate the accuracy of \sysname, we first randomly sampled 100 WeChat and 50 Baidu mini-programs that were identified as containing OBA misuse and manually inspected each case to check whether \sysname mistakenly detects OBA misuses in a mini-program while it does not (False Positives, FPs).
Additionally, we manually checked 100 WeChat and 50 Baidu mini-programs not flagged by \sysname to identify possible False Negatives (FNs).

\paragraphNew{False Positives} 
Among the WeChat samples, we identified 3 (3\%) FPs, all related to \textbf{M3} (plaintext transmission of sensitive data). 
These FPs arise from the lack of semantic context, where certain numeric or identifier-like fields coincidentally match the phone number pattern but are unrelated to authentication. 
In other cases, fields containing our test phone number were incorrectly flagged as sensitive, even though they had no impact on identity verification. 
We did not observe any FPs in Baidu mini-program samples.


\paragraphNew{False Negatives} Our manual inspection indicates that \sysname has no observed FNs among the mini-programs it successfully analyzed, meaning that all mini-programs containing OBA misuse were correctly identified.
We attribute this to three key factors aligned with our analysis pipeline: (1) The static pre-filter reliably captures mini-programs that incorporate OBA by parsing privacy declarations. Since developers often over-claim sensitive data usage in privacy agreements to comply with platform policies and legal requirements, this step ensures high recall during filtering.
(2) Our dynamic interaction model accurately executes OBA trigger paths by navigating to OBA-related pages and interacting with UI elements, ensuring that security-critical flows are actually invoked.
(3) The traffic monitor only inspects concrete runtime OBA messages, avoiding speculative assumptions and ensuring that actual misuse behaviors are captured.
We note that 13 mini-programs, which either use fully encrypted traffic or have access restrictions (allowing only a dedicated group of people, such as company employees, to log in), cannot be successfully tested end-to-end.
For such cases, we conservatively exclude them from both FN and FP counts.

\subsection{Validation of Detection Effectiveness}
\label{subsec:validation}

To validate the effectiveness of each core component, we conducted a module-wise evaluation on 100 randomly selected mini-programs across WeChat and Baidu.

\paragraphNew{Static Pre-filter} 
We manually verified that 26 WeChat and 27 Baidu mini-programs in the sampled set declared the use of OBA-based login mechanisms via platform-provided APIs. \sysname's static pre-filter successfully identified 25 WeChat mini-programs that fall within our measurement scope and all 27 Baidu mini-programs. The one missed WeChat case declares OBA in its metadata, but actually redirects users to another mini-program under the same developer for authentication via a plugin. As it bypasses the standard platform OBA flow, it is not subject to the misuse issues we studied. This confirms that the static pre-filter does not introduce FNs in detecting mainstream OBA usage.

\paragraphNew{Login Page Identifier}
Among the above 52 mini-programs with OBA, we find that 16 WeChat and 11 Baidu mini-programs are non-obfuscated. In these cases, \sysname successfully located the OBA trigger page and the corresponding UI element. For the remaining 25 obfuscated mini-programs, \sysname extracts all registered page routes and leverages an OCR-assisted module with a traversal algorithm that achieves both high accuracy and efficiency in identifying login interfaces. Across our full measurement, the OCR-Assisted Finder accurately locates OBA entry points in 133 obfuscated WeChat and 32 Baidu mini-programs, enabling misuse detection. These results demonstrate that combining route extraction and visual interface recognition is essential for achieving reliable coverage across both non-obfuscated and obfuscated mini-programs.

\paragraphNew{Dynamic Analyzer}
Through manual interaction, we confirm that 23/25 WeChat and 27/27 Baidu mini-programs successfully trigger the OBA flow and produce analyzable network traces. \sysname achieves identical results.
Among these, we manually verify that 4 WeChat mini-programs exhibited detectable misuses (one with \textbf{M1}, one with \textbf{M2}, and two with \textbf{M3}), while 2 Baidu mini-programs have the \textbf{M1} issue.
\sysname correctly flags all of them. For the other 2 WeChat mini-programs, we find that \sysname successfully locates and interacts with the OBA login page, but the expected OBA flow was not activated.
Upon manual investigation, these mini-programs either exceed their daily OBA usage limit, are for internal use only (e.g., using a whitelist-based phone number check), or redirect the OBA output to a custom identity verification flow.

\subsection{Comparison with Existing Approaches}

We compare \sysname with previous mini-program studies, from both vulnerability-scope and methodology perspectives. 

\begin{figure}[htbp]
	\centering
	\includegraphics[width=0.9\columnwidth]{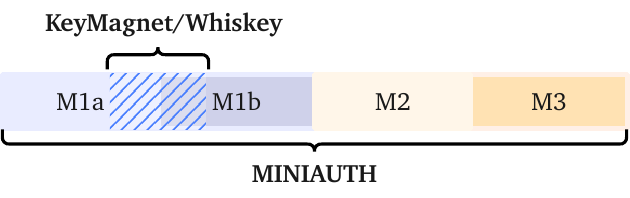}
	\caption{The vulnerability-scope Comparison}
	\label{fig:comparison_scope}
\end{figure}

KeyMagnet~\cite{DBLP:conf/ndss/ShiYZ0Y0025} and Whiskey~\cite{chen2025whiskey} overlap with our M1 findings, especially M1b.
However, KeyMagnet relies on static data-flow analysis and therefore, cannot handle obfuscation and dynamic loading.
Accordingly, it successfully analyzed only 6.2\% (2,748/44,273) of WeChat mini-programs in our dataset.
Additionally, due to the lack of runtime context, it cannot distinguish M1a from M1b, and cannot detect M2 and M3.
Lastly, it supports only mini-programs on WeChat, uncovering other platform, such as Baidu.
Whiskey is a concurrent dynamic credential-leakage detector, that can detect M1 by monitoring network traffic.
However, it also cannot distinguish M1a from M1b, as it does not consider the current authentication context when credentials are observed in the traffic.
Like KeyMagnet, Whiskey does not cover M2 or M3.
Table~\ref{fig:comparison_scope} illustrates the vulnerability-scope comparison between \sysname and existing work, while Table~\ref{tab:comparison} summarizes the comparison results on our dataset.
Because Whiskey's artifact is not publicly available, we are not able to compare \sysname with it.

\begin{table}[htbp]
\centering
\caption{Results of comparison with existing approaches.}
\label{tab:comparison}
\small
\begin{tabular}{lrr} 
\toprule
 & \textbf{MINIAUTH} & \textbf{KeyMagnet} \\
\midrule
WeChat Apps & 10,159 & 2,748  \\
\quad M1 / M2 / M3 & 1,201 / 257 / 282 & 490\textsuperscript{*} / 0 / 0 \\
Baidu Apps & 641 & 0 \\
\quad M1 / M2 / M3 & 51 / 2 / 41 & 0 / 0 / 0\\
\bottomrule
\multicolumn{3}{l}{*: Failed to distinguish M1 subcase (1) and (2).} \\ 
\end{tabular}
\end{table}



\subsection{Cross-platform Similarity and Divergence in OBA Misuse} 

We conduct a cross-platform analysis to understand whether OBA misuses in mini-programs are platform-specific or persist across different super app ecosystems. 
From our results, we identify 43 overlapping mini-programs between WeChat and Baidu by matching identical or similar names. Among these, all 43 also appear on WeCom~\cite{wecom-mp}, while 17 and 8 corresponding versions are found on Alipay~\cite{alipay-mp} and TikTok~\cite{tiktok-mp}, respectively.
\Cref{tab:applist} in Appendix summarizes our testing results. 
Despite differences in super app environments, OBA misuses are highly consistent: 43/43 WeCom, 42/43 Baidu, 17/17 Alipay, and 8/8 TikTok mini-programs show nearly identical issues as their WeChat counterparts, with similar API flows and traffic structures. 
This indicates that once a mini-program is insecure on one platform, its counterparts on other platforms are very likely to share the same weakness.

Additionally, we discovered cases where the same mini-programs adopt different security practices across platforms and highlight the following insights:

\noindent \textbf{I. The weakest link across multiple platforms determines the security of a mini-program.} 
We observed significant security disparities where mini-programs deployed robust measures on one platform but retained insecure implementations on another. For instance, several popular services utilized encrypted traffic for their WeChat versions to secure the OBA process, while their Baidu counterparts exposed \texttt{session\_keys} in plaintext. Since these versions typically share a unified back-end, an exploit on the weaker platform can compromise user accounts across both ecosystems.


\noindent \textbf{II. Financial cost might be a factor influencing the choice of authentication method.}  
In some cases, the same mini-program adopts divergent authentication mechanisms across platforms. For example, an online education provider, Zhihuishu~\cite{zhs-mp}, utilizes standard OBA on Baidu but requires manual phone number entry and third-party SMS verification on WeChat. Such discrepancies are often driven by platform-specific policies, such as the per-request costs associated with enabling OBA features on WeChat~\cite{wx-phonenumber-price}. 
This lack of uniformity weakens security and may push users toward clunkier, less protected authentication options.

\noindent \textbf{III. Mandatory circumventing platform checks might not achieve what is expected.}  
Instead of improving their implementations, some developers deliberately modified parameter names to evade the built-in security checks of super apps, such as the hard-coded detection of \texttt{session\_key} in WeChat. For instance, we identified a mini-program that utilized \texttt{session\_key} in its Baidu version but replaced it with \texttt{sk} in the WeChat version while preserving identical semantics. This adversarial adaptation highlights the inherent limitations of static analysis and underscores the necessity of dynamic solutions for robust OBA security auditing.
These findings suggest that platform heterogeneity and developer adaptation strategies can lead to inconsistent security postures, even within the same service. They also emphasize the need for cross-platform security auditing and consistent enforcement policies to mitigate such fragmented protections.


\subsection{Real-world Case Study}

In this section, we present a representative OBA misuse case in real-world mini-programs to showcase how to exploit these misuses and their impact. Also, we present another misuse case in the Appendix to illustrate cross-platform threats.

\begin{figure}[t]
	\centering
	\includegraphics[width=0.9\columnwidth]{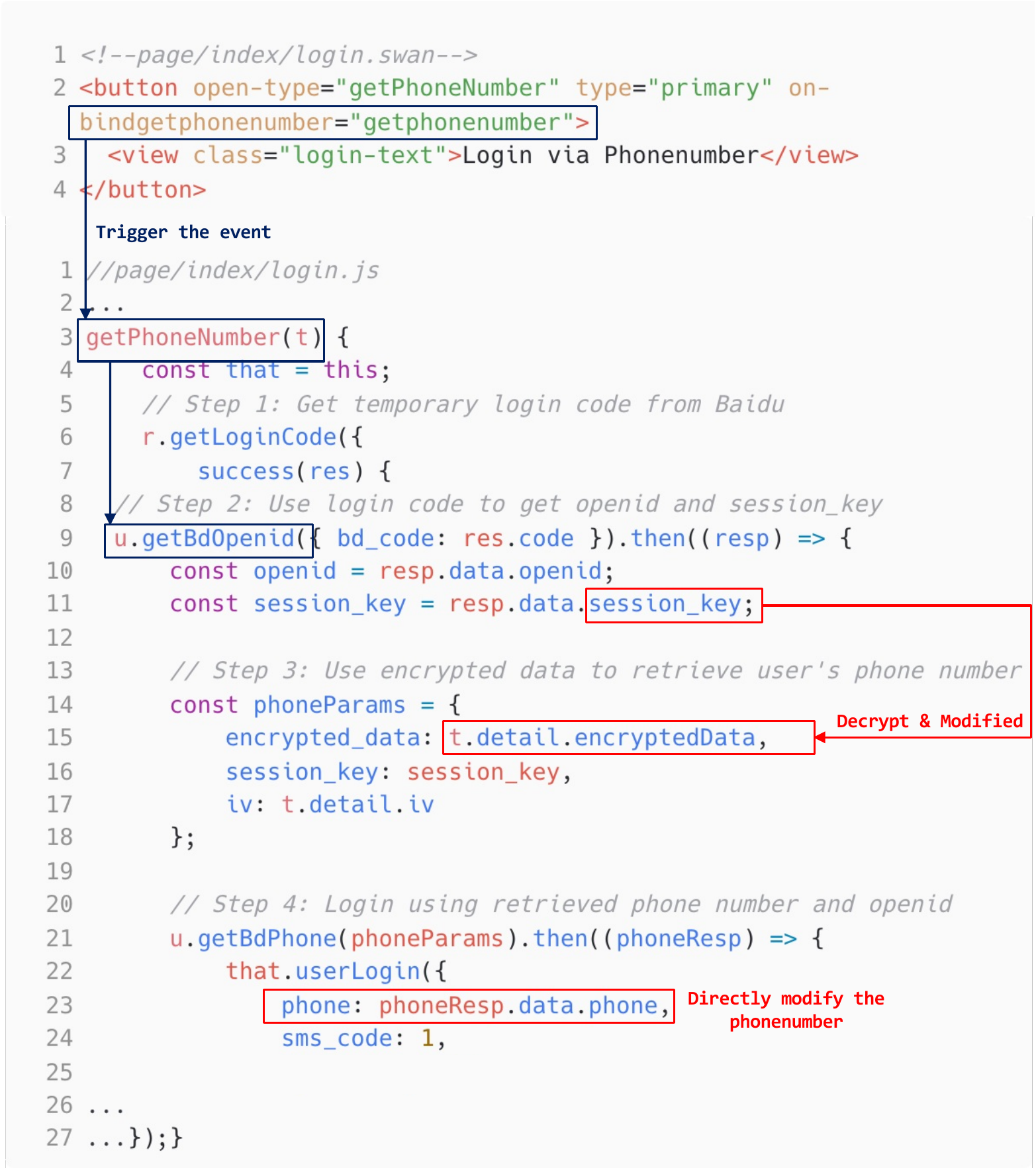}
	\caption{The code snippet of ADJ's Baidu mini-program.}
	\label{fig:aidaijia_code}
\end{figure}

\paragraphNew{ADJ Mini-program} ADJ is a popular ride-sharing mini-program that allows users to request professional drivers through their phones. It ranks among the top three in China's driver-for-hire service market, with over 100,000 registered drivers. ADJ is available under the same name on both the Baidu and WeChat platforms. \Cref{fig:aidaijia_code} shows the core logic for retrieving a user's phone number in the Baidu version. When the user taps the login button on the login page, it triggers the \texttt{getPhoneNumber} event (Line 2). Inside this event, the program first calls \texttt{getLoginCode} to obtain a temporary OBA \texttt{code} (Line 9). It invokes \texttt{getBdOpenid} but leaks the \texttt{session\_key} in the response.

\begin{figure}[t]
	\centering
	\includegraphics[width=1\columnwidth]{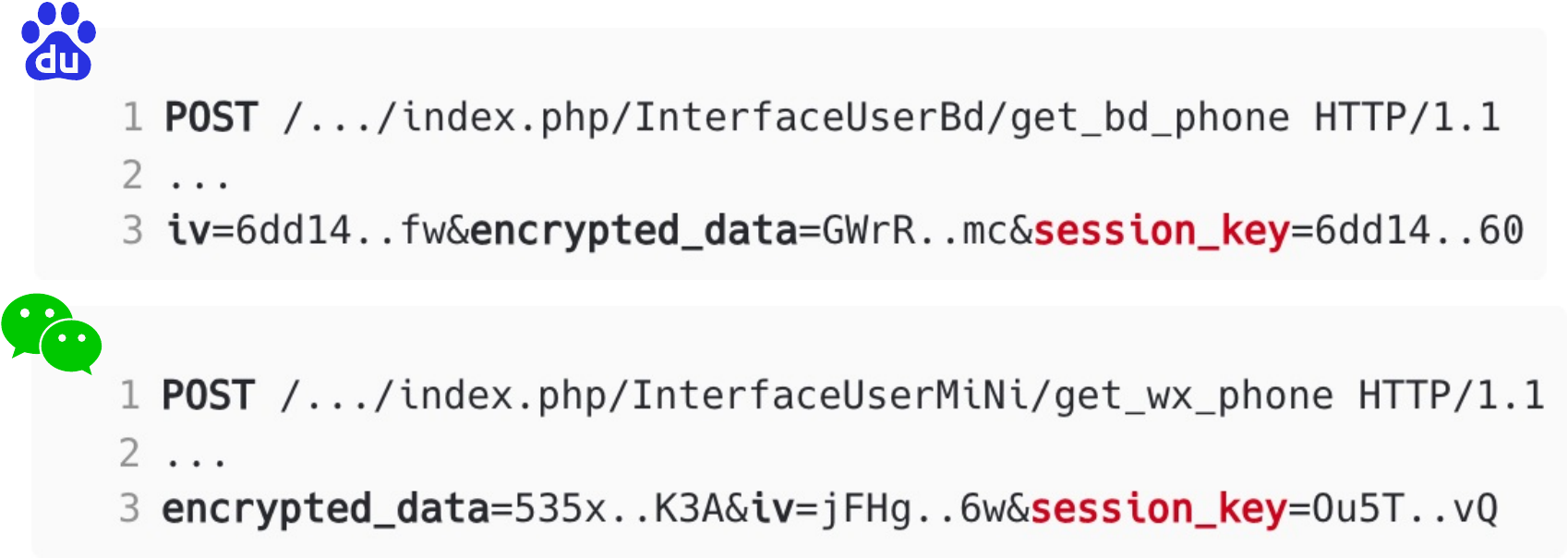}
    \caption{Traffic snippet of \texttt{getPhoneNumber} in the ADJ mini-program.}
	\label{fig:aidaijia_capture}
\end{figure}
Subsequently, the mini-program forwards \texttt{encryptedData}, \texttt{IV}, and \texttt{session\_key} to the \texttt{getBdPhone} API (Line 21), which decrypts the phone number.
The decrypted plaintext phone number is then forwarded to the \texttt{userLogin} function (Line 22) to complete the authentication process.
This design leads to two critical misuses: 1) \textbf{M1a}, involving insecure handling of encrypted data alongside a leaked \texttt{session\_key}, and 2) \textbf{M3}, involving the reuse of plaintext phone numbers without server-side validation.

An attacker can exploit the misuses in two ways to log in to this mini-program using the victim's identity when the attacker runs the mini-program on her device: (1) exploit \textbf{M1} to create \texttt{encryptedData} with the victim's phone number; (2) exploit \textbf{M3} by directly replacing the plaintext phone number in the final login request with the victim's.
Since ADJ is a ride-hailing and designated driving service, this vulnerability means that an attacker who knows a victim’s phone number could hijack the victim’s account, access sensitive booking history, and even place fraudulent orders under the victim’s identity, posing serious threats to both personal privacy and financial security. Given ADJ's massive user base and integration with critical features like identity verification and payments, authentication flaws pose substantial security risks to millions of users.

Furthermore, we observed that ADJ's WeChat version is heavily obfuscated, which prevents static-analysis-based existing approaches from identifying security issues.
Nevertheless, \sysname can successfully trigger the OBA workflow and reveal the misuses.
\Cref{fig:aidaijia_capture} compares the traffic context of \texttt{getPhoneNumber} from both platforms. Aside from slight differences in request routing endpoints, the request bodies and exposed sensitive fields were nearly identical.
This consistency suggests that developers often reuse business logic and OBA integration code across platforms, which can lead to systematic replication of security flaws.

%% file: 8.Discussion.tex
\section{Discussion and Limitation}
\label{sec:discussion}

\paragraphNew{Real-world Impact Estimation}
Mini-program platforms do not disclose per-program downloads, active users, or installation counts, making it difficult to directly measure the popularity of affected mini-programs. Prior work used user ratings and rating counts as popularity indicators~\cite{yang2022cross}, but such signals are optional and can be sparse or biased. To complement these indicators, we measured passive DNS (pDNS) activity for the domains associated with the representative cases in \Cref{tab:disclosure_cases} and report the corresponding monthly visit counts in the Appendix. pDNS traffic is not equivalent to user counts and only captures the visible portion of domain-level activity; therefore, the reported monthly visits should be interpreted as a conservative lower-bound estimate rather than a complete measurement of mini-program usage.

As shown in \Cref{tab:disclosure_cases}, these OBA misuses affect applications across critical sectors, including banking, healthcare, logistics, public services, and IoT device management. Even under this lower-bound measurement, several affected services show substantial traffic: DeliveryApp-Z is associated with more than 20 million monthly visits, MuseumApp-B with more than 11 million, and BankingApp-X with more than 7 million. These results indicate that the discovered vulnerabilities can affect widely used real-world services, while smaller pDNS counts should not be interpreted as evidence of negligible impact.

\paragraphNew{Root Causes} 
We summarize the root causes of OBA misuse in mini-programs into the following three aspects:
\textit{(1) Unclear separation of front-end and back-end roles.}
Developers often misunderstand which parts of the OBA process must remain server-side. In particular, \texttt{session\_key} is a back-end credential used to decrypt and verify platform-provided data, but some implementations expose it to the untrusted mini-program front-end or return it in later network responses. This breaks the intended credential-isolation boundary and leads to M1.

\textit{(2) Lack of authenticity and freshness verification on the back-end.}
Many mini-programs treat decrypted data or platform-issued identifiers as proof of identity by themselves. This ``decryption-equals-authentication'' fallacy is unsafe because the front-end is attacker-controlled under our threat model. Back-end logic must bind \texttt{OpenID}, \texttt{UnionID}, phone numbers, or other sensitive attributes to a fresh platform-verified OBA exchange; otherwise, attackers can replay static identifiers or modify plaintext identity fields, leading to M2 and M3.

\textit{(3) Error-prone platform abstraction and limited enforcement.}
Mini-program OBA APIs are designed for developer usability, but their compact abstraction can obscure the trust boundary among the front-end, the developer back-end, and the super app back-end. Identity fields and cryptographic material are exposed through closely related interfaces, while platform-encrypted user attributes are delivered through the front-end before reaching the developer server. Current platform-side checks rely mainly on static source scanning and cannot reliably detect runtime misuse in network traffic, especially when developers rename sensitive fields.

Overall, these issues reflect a dual failure: platform abstractions make OBA look like ordinary data plumbing, while developer implementations often omit the back-end verification needed to preserve authenticity, freshness, and server-only cryptographic state.

\paragraphNew{Lessons Learned}
Based on our root cause analysis, we provide several practical takeaways for both mini-program developers and platform vendors to avoid OBA misuse.

Mini-program developers should avoid returning credentials, such as \texttt{session\_key}, to the mini-program. These fields must be processed strictly on the server side. 
Additionally, combining multiple identities for authentication, such as \texttt{UnionID} and \texttt{OpenID}, can enhance identity binding and reduce the risk of impersonation.
Lastly, additional measures to secure communication between the mini-program and its server, such as applying digital signatures and encryption, can further protect the authentication process.


From the mini-program platform side, improving the design of OBA interfaces, such as using different interfaces to deliver identities and decryption keys, can reduce the risk of misuse.
Furthermore, platforms could perform runtime analysis before publishing the mini-programs in addition to the current pure static mechanisms.
Lastly, common cryptographic issues, such as using part of the key as the IV, could be mitigated by enforcing stricter platform policies. 


\paragraphNew{Limitations} While \sysname demonstrates practical effectiveness, it has two main limitations. First, its dynamic analyzer depends on successful UI rendering and stable network connectivity. In some cases, mini-programs may detect the presence of an MITM-based HTTP proxy, which is used by \sysname to monitor traffic. When such detection occurs, the mini-program may refuse to load content or terminate the authentication flow, preventing further analysis.
Second, for enterprise or internal-only mini-programs that adopt an extra layer of traffic encryption, \sysname is unable to extract meaningful information from the traffic.
However, such cases remain difficult even for manual analysis, posing an open challenge for future mini-program security research.

%% file: 9.RelatedWork.tex
 \section{Related Work} 
\label{sec:related_work}

\paragraphNew{Mini-program Security} Research on security and privacy within mini-program ecosystems has significantly expanded.
Zhang et al.~\cite{zhang2021measurement} introduced \textit{MiniCrawler} to analyze API usage and obfuscation, while Lu et al.~\cite{lu2020demystifying} studied resource management flaws. 
Wang et al.~\cite{wang2022characterizing} proposed \textit{WeDetector} to detect common bugs. 
Several works addressed authentication issues. Zhang et al.~\cite{zhang2022identity} revealed identity confusion in super apps, Yang et al.~\cite{yang2022cross} developed \textit{CMRFScanner} for cross-request forgery, Zhang et al.~\cite{DBLP:conf/ccs/ZhangHYDGLGD24} proposed \textit{MiniCAT} for cross-page forgery, and Shi et al.~\cite{DBLP:conf/ndss/ShiYZ0Y0025} presented \textit{KeyMagnet} to detect credential leaks. 
Other studies examined malicious or privacy-violating behaviors, including malware datasets~\cite{DBLP:conf/ndss/00030025}, taint-based tools~\cite{wang2023taintmini,DBLP:conf/sats/YuKC23}, and hybrid analysis for privacy inconsistencies~\cite{DBLP:journals/tosem/WangLWL0LW24}. 
\textit{MiniChecker}~\cite{wang2024minichecker} further identified systemic flaws in permission pop-ups. 
Cross-platform inconsistencies have also been explored, e.g., \textit{APIDIFF}~\cite{wang2023one}, \textit{MiniTracker}~\cite{li2023minitracker}, and recent work on protocol flaws and key misuse~\cite{DBLP:conf/raid/BaskaranZMY23,zhang2023don}. 
Other directions include third-party libraries~\cite{DBLP:conf/sats/TaoS0WLL23}, data minimization~\cite{DBLP:conf/sats/WangZWW23}, dark patterns~\cite{DBLP:conf/sats/LongXWON23}, domain whitelist flaws~\cite{DBLP:conf/sats/ZhangZL000023}, plugin signatures~\cite{DBLP:conf/sats/ZhaoZW23}, and cross-user leakage~\cite{xpo-mp}. 

In contrast, our work addresses the systematic detection of mini-program OAuth-based Authentication (OBA) misuse. We do not claim that every underlying bug pattern is entirely new: credential exposure in M1b overlaps with KeyMagnet and other credential-leakage studies. Our novelty lies in distinguishing the timing and exploitability of M1a and M1b, covering static-identifier and plaintext-identifier misuse (M2/M3), and dynamically validating these patterns across obfuscated mini-programs and multiple platforms.


\paragraphNew{OAuth Security}
OAuth is a widely adopted protocol for authorization and single sign-on (SSO), and its security has been extensively studied. Early efforts analyzed vulnerabilities in OAuth and SSO implementations across web and mobile platforms. For instance, Sun et al.~\cite{DBLP:conf/ccs/SunB12} and Wang et al.~\cite{DBLP:conf/ccs/0010XWC13} revealed severe flaws in OAuth deployments among popular services, while Ghasemisharif et al.~\cite{DBLP:conf/uss/GhasemisharifRC18, DBLP:conf/sp/GhasemisharifKP22} systematically investigated SSO risks in account and session management. Fett et al.~\cite{DBLP:conf/ccs/FettKS16, DBLP:conf/sp/FettHK19} provided the first formal security analysis of OAuth 2.0 and its extensions. Symbolic and static analysis tools like S3KVetter~\cite{DBLP:conf/uss/YangLCZ18}, Cerberus~\cite{DBLP:conf/ccs/Rahat0T22}, and AuthSaber~\cite{DBLP:conf/ccs/Rahat0024} were proposed to detect logic flaws and verify real-world OpenID Connect implementations. Cao et al.~\cite{DBLP:conf/uss/CaoMSF24} mitigated token overprivilege in OAuth. Subsequent studies explored how authentication states affect vulnerability discovery~\cite{DBLP:conf/sp/RautenstrauchMHHS24} and examined security risks arising from the interplay between local and federated login paths~\cite{DBLP:conf/sp/GhasemisharifKP22}.

Unlike standard OAuth/OIDC, mini-program OBA lacks browser-visible redirects and signed identity tokens, and operates within closed, API-dependent environments. Prior web SSO scanners can perform GUI-based exploration, but they rely on browser-visible states such as DOM elements, URLs, redirects, and protocol parameters. Mini-program OBA exposes the relevant authentication states through JS Bridge APIs, native dialogs, and proprietary vendor exchanges instead. \sysname is tailored to this ecosystem by combining GUI/OCR-assisted triggering with OBA-specific runtime monitoring.



%% file: 10.Conclusion.tex
\section{Conclusion}
\label{sec:conclusion}
In this work, we conducted the first large-scale investigation of OBA misuses within mini-program ecosystems and developed \sysname, a dynamic analysis framework designed to uncover runtime logic flaws that evade static detection. By evaluating over 46K WeChat and Baidu mini-programs, we demonstrated that OBA misuses are widespread, affecting 13.8\% of WeChat and 45.4\% of Baidu mini-programs. Our analysis systematized these vulnerabilities into critical classes, including client-side identity construction, unauthorized credential exposure, and reliance on static or plaintext identifiers. Notably, we uncovered platform-level design flaws, such as predictable IV reuse in Baidu, which facilitate practical brute force attacks. Our cross-platform case studies further revealed that insecure implementation patterns are frequently replicated across ecosystems due to code reuse and adversarial adaptation to platform-specific security checks. These findings underscore the urgent need for platform vendors to refine OBA API designs and for developers to move beyond superficial compliance toward robust, cryptographically secure authentication practices.

%% file: Ethical_OpenScience.tex
\appendix

\section*{Ethics Considerations}
\label{sec:ethical}
In conducting this work, we strictly adhered to ethical guidelines and performed a multi-party stakeholder analysis to balance research transparency with user safety.

\paragraphNew{Controlled Testing and Safeguards} 
All exploit validations were restricted to accounts, devices, and network environments fully controlled by the research team. We did not target third-party user accounts or interfere with live production services. Test traffic was rate-limited to 20 requests per minute, replayed only within a closed lab network, and all logs containing sensitive patterns were stored on isolated machines with restricted access. 

\paragraphNew{Responsible Disclosure and Follow-up}
We conducted all experiments ethically and disclosed our findings to relevant stakeholders through multiple channels. We first reported the platform-level cryptographic flaw, where the IV was derived from key material, and third-party OBA misuses (M1--M3) to vendors, including Baidu and Tencent.
They confirmed the platform flaw and acknowledged the misuse reports, while noting that third-party mini-programs were outside their direct response scope.
We then reported all identified OBA misuse cases to national coordinators, including CNVD~\cite{CNVD}, CNNVD~\cite{CNNVD}, and CNCERT/CC~\cite{CNCERT/CC}, and actively coordinated with CNCERT/CC.
During this process, the national CERT updated its review criteria to classify third-party mini-programs as ``non-critical assets'', leaving many reports under extended review despite initial acknowledgment.
In parallel, we batch-reported affected cases to Tencent and Baidu SRC programs, which acknowledged receipt and stated that follow-up work would be conducted, although both programs currently accept reports only for their official first-party applications.
We also tried to reach third-party developers directly by extracting 556 reachable email addresses from regulatory metadata and emailing them with vulnerability summaries and repair suggestions.
Unfortunately, we did not receive any response from these developers.
Nevertheless, to assess real-world remediation, after 90 days we followed up on 11 confirmed vulnerable cases (\Cref{tab:disclosure_cases} in the Appendix), spanning high-impact services such as banking, healthcare, smart locks, and gas payments.
Fortunately, we found that several mini-programs have effective fixes deployed, including traffic encryption, removing \texttt{session\_key} from payloads, redesigning login flows, adding custom validations such as tokens or signatures, and introducing secondary authentication.
However, some other mini-programs remained unpatched, were removed, or received only superficial fixes, such as replacing OBA with a weak combination of \texttt{OpenID} and phone numbers. 

\paragraphNew{Ongoing Mitigation and Stakeholder Protection} 
To protect end users and developers from potential copycat exploits, we are implementing a 90-day embargo on the complete \sysname artifact following our latest round of outreach. The Appendix of this paper and our artifact repository include a vetted self-check guide to help developers detect and remediate M1--M3 patterns. We exclude the raw 130\,GB mini-program corpus from the public release to protect proprietary code and prevent the exposure of services that remain unpatched.

\section*{Open Science}
\label{sec:openscience}
In compliance with open science policies, we provide unobfuscated and documented source code for the \sysname framework together with an anonymized dataset of detection samples under an open-source license at \url{https://anonymous.4open.science/r/MiniAuth-83A3}. As noted in the Ethics section, we intentionally exclude the original mini-program corpus, real vulnerable identifiers, and direct exploit-enabling artifacts to protect developers and users. These redactions do not affect reproducibility of the analysis workflow: the repository includes the framework source, configuration templates, sanitized samples, and instructions needed to reproduce the detection pipeline and build upon our dynamic analysis framework.

\section*{Acknowledgements}
This paper was edited for grammar using Gemini. We thank the anonymous reviewers for their valuable comments and suggestions. This work was supported in part by the Natural Sciences and Engineering Research Council of Canada (NSERC) under Grant RGPIN-2025-03988, and through a Discovery Launch Supplement (DGECR-2025-00455) and a DND/NSERC Discovery Grant Supplement (DGDND-2025-03988).
Any opinions, findings, and conclusions in this paper are those of the authors and do not necessarily reflect the views of NSERC or DND. The authors from Shandong University were supported by the Natural Science Foundation of Shandong Province (Grant No. ZR2025MS991).

%% file: Appendix.tex
\clearpage
\begin{figure*}[b]  
	\centering
	\includegraphics[width=0.9\textwidth]{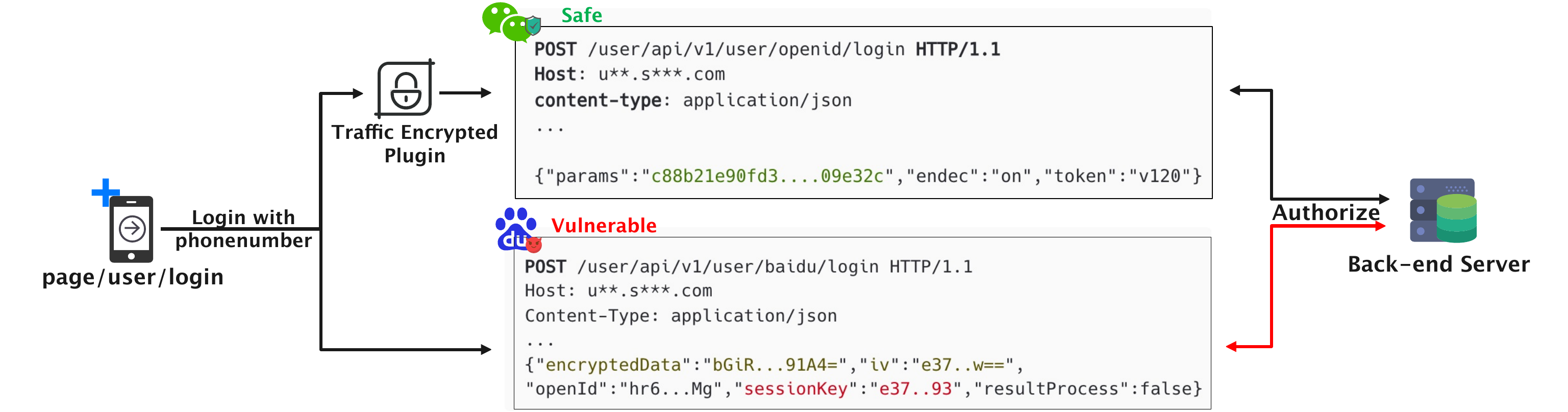}
	\caption{The workflow and traffic snippets of the SYDEY Hospital mini-program.}
	\label{fig:sydeyCase}
\end{figure*}

\appendix

\section{Appendix}
\label{sec:Appendix}

\begin{figure}[htbp] 
\centering 
\includegraphics[width=1\columnwidth]{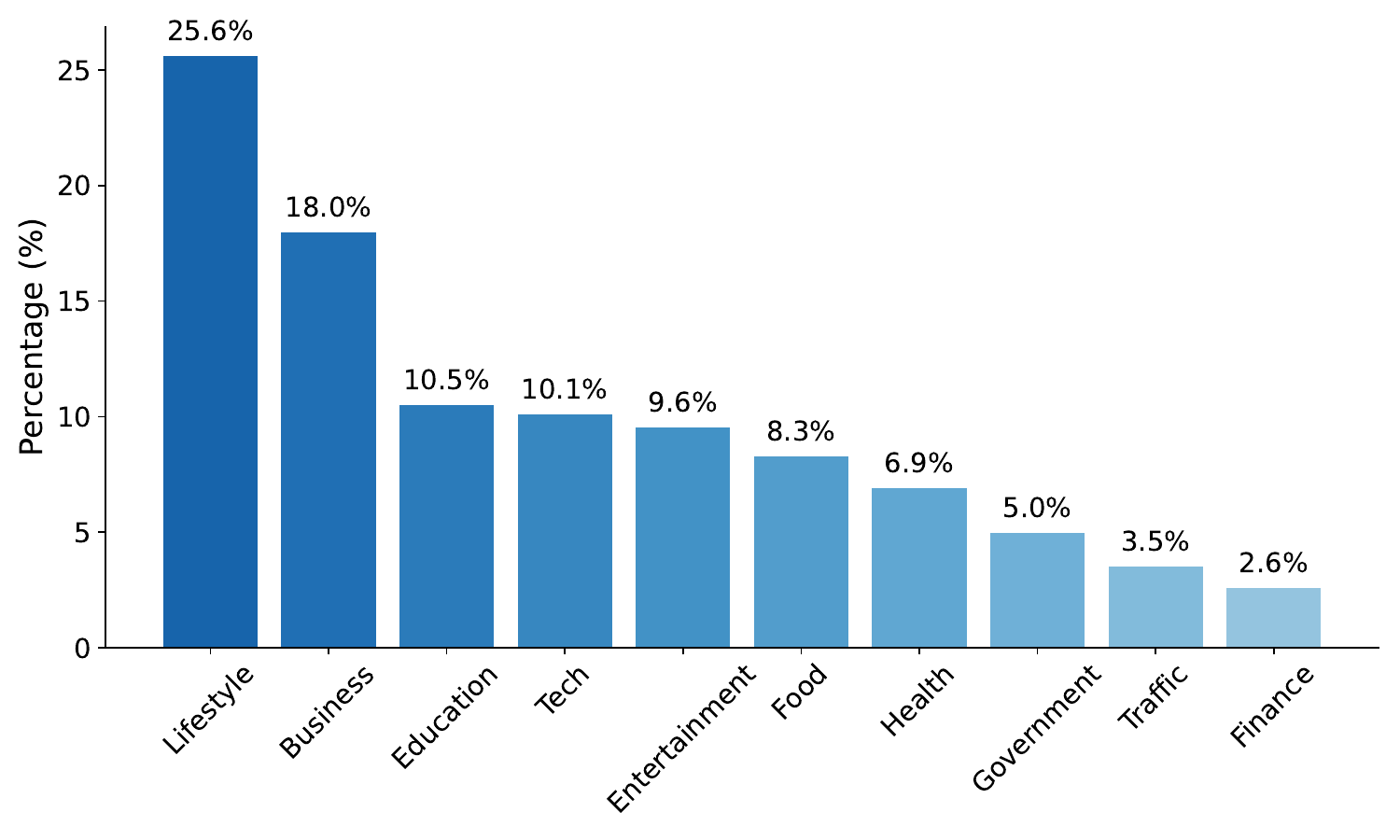} \caption{The OBA misuse statistic based on categories.} \label{fig:cate_statis} 
\end{figure}

\paragraphNew{Measurement Results by Category} 
We further categorize the affected mini-programs to measure the impact of OBA misuse across different application domains. As shown in~\Cref{fig:cate_statis}, the \textit{Lifestyle} category has the highest proportion of misuse (25.6\%). Notably, the \textit{Business} category (18.0\%) and the \textit{Government} category (5.0\%) also show unexpected misuse rates. These mini-programs typically require frequent access to sensitive personal information (e.g., national ID numbers, payment credentials, or biometric data), making secure OBA implementation particularly critical in these domains.

\paragraphNew{Cross-platform Measurement Results} 
\Cref{tab:applist} lists all OBA misuse cases across multiple platforms.  

\paragraphNew{Follow-up Disclosure Investigation}
\Cref{tab:disclosure_cases} summarizes representative cases with their primary functionality, potential risks, response status, and fixed solutions.

\paragraphNew{Case Study: SYDEY Hospital Mini-program} Here we present another misuse case in the cross-platform scenario. SYDEY Hospital is a healthcare mini-program that provides services such as appointment booking, health report retrieval, and medical insurance payment. Both its WeChat and Baidu versions are heavily obfuscated, preventing existing static analysis.

As illustrated in \Cref{fig:sydeyCase}, the WeChat version employs a custom third-party plugin to facilitate secure OBA login. Through traffic inspection, we observed that all sensitive exchanges, including those for retrieving phone numbers, are encrypted at the transport layer or wrapped within a secure plugin interface.
This effectively prevents attackers from sniffing or modifying session-related credentials.In contrast, the Baidu version lacks such traffic protection and has the \textbf{M1} issue.
Sensitive data, including the \texttt{session\_key}, is transmitted in plaintext and can be intercepted during login.
Since both mini-programs interact with the same back-end server (they access the same API endpoints with the same payload structure), an attacker can exploit the vulnerability in the Baidu version to forge credentials and gain unauthorized access to the unified back-end, even if the WeChat version appears secure.

This case reinforces our cross-platform finding: securing one platform variant is insufficient when all versions share the same back-end. A weakness in the Baidu deployment compromises the overall system, underscoring the need for consistent protection across all platforms.

\begin{table*}[htbp]
\centering
\caption{Summary of disclosed mini-program cases and follow-up status.}
\label{tab:disclosure_cases}
\small
\begin{tabular}{p{2.0cm}p{2.7cm}p{3.4cm}p{1.9cm}p{2.9cm}r}
\toprule
Mini-program Alias & Primary Function & Potential Risk & Follow-up Status & Fix or Remaining Risk & Monthly Visits \\
\midrule
MuseumApp-A & Museum ticket booking & Unauthorized access to ticket information, including national ID and phone number & Fixed & Encrypted network communications & 3,602 \\
\midrule
MuseumApp-B & Museum ticket management & Unauthorized access to ticket information, including national ID and phone number & Fixed & Encrypted network communications & 11,660,174 \\
\midrule
MuseumApp-C & Museum ticket booking & Unauthorized access to sensitive ticket information & Unpatched & \textcolor{red}{No fix observed at follow-up} & 742,834 \\
\midrule
BankingApp-X & Digital banking and account overview & Partial access to masked account data& Fixed & Replaced vulnerable flow with custom SMS-based authentication & 7,147,419 \\
\midrule
SmartLockApp & IoT-based smart lock management & Device takeover via victim's account & Removed / unknown & \cellcolor{yellow!25}No patched version found; possibly unpublished or renamed & 6,484 \\
\midrule
ETCApp & Vehicle toll and license management & Unauthorized access to vehicle and personal details & Weakly fixed & \cellcolor{yellow!25}Patched, but still relies on \texttt{OpenID} and phone number for login & 1,101,970 \\
\midrule
DeliveryApp-Z & On-demand delivery and logistics & Access to order history, addresses, and personal data & Fixed & Removed \texttt{session\_key} from request payload & 20,397,517 \\
\midrule
Floodgate-Control App & BLE-based floodgate control system & Full control over floodgate devices & Fixed & Added custom OBA validation with user whitelist & 166,019 \\
\midrule
HealthMonitorApp & Diabetes and health tracking & Unauthorized access to sensitive health metrics (height, weight, blood glucose) & Removed / unknown & \cellcolor{yellow!25}No patched version found; possibly unpublished or renamed & 19,581 \\
\midrule
GasStationApp & Gas station self-service and payments & Execute fuel payments using victim's account & Fixed & Added custom token and signature validation & 14,636 \\
\midrule
PostalServiceApp & Provincial-level postal service platform & Access to sensitive user and order information & Fixed & Added timestamp and signature validation to prevent request forgery & 39,747 \\
\bottomrule
\end{tabular}
\end{table*}

\begin{table*}[!htp]
 \centering
\caption{Cases of OBA misuses in mini-programs existing on multiple platforms.}
\label{tab:applist}

\begin{tabular}{|c|c|c|c|c|c|}
\hline
\rowcolor[HTML]{FFFFFF} 
\textbf{Name of Mini-Program} & \textbf{WeChat} & \textbf{WeCom} & \textbf{Baidu} & \textbf{Alipay} & \textbf{TikTok} \\ \hline
\rowcolor[HTML]{C0C0C0} 
SYDEY Hospital         & \ding{109} & \ding{109} & \ding{108} & \ding{108} &   \\ \hline
AD Ride-hailing        & \ding{108} & \ding{108} & \ding{108} & \ding{108} & \ding{108} \\ \hline
\rowcolor[HTML]{C0C0C0} 
ACW Lottery            & \ding{108} & \ding{108} & \ding{108} &           &           \\ \hline
CW Insurance Agent     & \ding{108} & \ding{108} & \ding{108} &           &           \\ \hline
\rowcolor[HTML]{C0C0C0} 
FBL Legal Service      & \ding{108} & \ding{108} & \ding{108} & \ding{108} &           \\ \hline
XXJ Script Game        & \ding{108} & \ding{108} & \ding{108} & \ding{108} &           \\ \hline
\rowcolor[HTML]{C0C0C0} 
JT Express             & \ding{108} & \ding{108} & \ding{108} & \ding{108} & \ding{108} \\ \hline
QZ Exhibition          & \ding{108} & \ding{108} & \ding{108} & \ding{108} &           \\ \hline
\rowcolor[HTML]{C0C0C0} 
EDJ Ride-hailing       & \ding{108} & \ding{108} & \ding{108} & \ding{108} & \ding{108} \\ \hline
ZP Housing             & \ding{108} & \ding{108} & \ding{108} &           & \ding{108} \\ \hline
\rowcolor[HTML]{C0C0C0} 
ZQ Research            & \ding{108} & \ding{108} & \ding{108} &           &           \\ \hline
TH Testing             & \ding{108} & \ding{108} & \ding{108} &           &           \\ \hline
\rowcolor[HTML]{C0C0C0} 
ZBS Trademark Market   & \ding{108} & \ding{108} & \ding{108} &           &           \\ \hline
DLYZ E-gov Service     & \ding{108} & \ding{108} & \ding{108} & \ding{108} &           \\ \hline
\rowcolor[HTML]{C0C0C0} 
QMKX Quick Repair      & \ding{108} & \ding{108} & \ding{108} & \ding{108} &           \\ \hline
BBSZ Trademark Transfer& \ding{108} & \ding{108} & \ding{108} & \ding{108} &           \\ \hline
\rowcolor[HTML]{C0C0C0} 
SS Courier             & \ding{108} & \ding{108} & \ding{108} & \ding{108} &           \\ \hline
HPG Screen Repair      & \ding{108} & \ding{108} & \ding{108} & \ding{108} &           \\ \hline
\rowcolor[HTML]{C0C0C0} 
YAB Insurance          & \ding{108} & \ding{108} & \ding{108} &           &           \\ \hline
TRJ BizPlan Service    & \ding{108} & \ding{108} & \ding{108} &           &           \\ \hline
\rowcolor[HTML]{C0C0C0} 
XCFZ Apparel           & \ding{108} & \ding{108} & \ding{108} &           &           \\ \hline
SXX Home Repair        & \ding{108} & \ding{108} & \ding{108} & \ding{108} &           \\ \hline
\rowcolor[HTML]{C0C0C0} 
HBPXW Training         & \ding{108} & \ding{108} & \ding{108} &           &           \\ \hline
QFL Legal Q\&A         & \ding{108} & \ding{108} & \ding{108} & \ding{108} & \ding{108} \\ \hline
\rowcolor[HTML]{C0C0C0} 
XFL RealEstate         & \ding{108} & \ding{108} & \ding{108} &           & \ding{108} \\ \hline
XHB Insurance          & \ding{108} & \ding{108} & \ding{108} &           &           \\ \hline
\rowcolor[HTML]{C0C0C0} 
JMZ Housekeeping       & \ding{108} & \ding{108} & \ding{108} & \ding{108} & \ding{108} \\ \hline
LKW CreditCard         & \ding{108} & \ding{108} & \ding{108} &           &           \\ \hline
\rowcolor[HTML]{C0C0C0} 
YQXXW Instrument Info  & \ding{108} & \ding{108} & \ding{108} & \ding{108} &           \\ \hline
TTS Recycling          & \ding{108} & \ding{108} & \ding{108} & \ding{108} &           \\ \hline
\rowcolor[HTML]{C0C0C0} 
BLKW Legal Q\&A        & \ding{108} & \ding{108} & \ding{108} &           &           \\ \hline
LPW Housing            & \ding{108} & \ding{108} & \ding{108} &           & \ding{108} \\ \hline
\rowcolor[HTML]{C0C0C0} 
DevCommunity           & \ding{108} & \ding{108} & \ding{108} &           &           \\ \hline
HSL Auto               & \ding{108} & \ding{108} & \ding{108} &           &           \\ \hline
\rowcolor[HTML]{C0C0C0} 
HQCG Immigration       & \ding{108} & \ding{108} & \ding{108} &           &           \\ \hline
YYYS Wellness          & \ding{108} & \ding{108} & \ding{108} &           &           \\ \hline
\rowcolor[HTML]{C0C0C0} 
XH Dictionary          & \ding{108} & \ding{108} & \ding{108} &           &           \\ \hline
\end{tabular}

\vspace{4pt}
\begin{center}
  \small \ding{108}: Found \& confirmed vulnerable; \ding{109}: Found but not vulnerable.
\end{center}
\end{table*}